\documentclass[10pt,journal,compsoc]{IEEEtran}
\ifCLASSOPTIONcompsoc
  \usepackage[nocompress]{cite}
\else
  \usepackage{cite}
\fi
\ifCLASSINFOpdf
\else
\fi
\usepackage{amsfonts}    
\usepackage{color}
\usepackage{graphicx}
\usepackage[dvips]{epsfig}
\usepackage{graphics}
\usepackage{arydshln}
\usepackage{amsmath} 
\usepackage{amssymb}  
\usepackage{subfigure}
\usepackage{subfloat}
\usepackage{caption}
\usepackage{algorithm}
\usepackage{algorithmic}

\def\bR{\mathbb{R}}

\def\cA{\mathcal{A}}

\def\cU{\mathcal{U}}

\def \qed {\hfill \vrule height6pt width 6pt depth 0pt}
\def\bee{\begin{equation}}
\def\ene{\end{equation}}
\def\been{\begin{equation*}}
\def\enen{\end{equation*}}
\def\beq{\begin{eqnarray}}
\def\enq{\end{eqnarray}}

\newtheorem{pro}{Proposition}[section]
\newtheorem{lem}{Lemma}[section]

\newtheorem{defi}{Definition}[section]

\begin{document}
%
\title{Dynamic Pricing and Mean Field Analysis for Controlling Age of Information}
%
%
%
%

\author{Xuehe~Wang,~\IEEEmembership{Member,~IEEE,}
        and~Lingjie~Duan,~\IEEEmembership{Senior~Member,~IEEE}
\IEEEcompsocitemizethanks{\IEEEcompsocthanksitem X. Wang is with the School of Artificial Intelligence, Sun Yat-sen University, China (E-mail: xwang21@e.ntu.edu.sg).
\IEEEcompsocthanksitem L. Duan is with the Pillar of Engineering Systems and Design, Singapore University of Technology and Design, Singapore (E-mail: lingjie\_duan@sutd.edu.sg).
\IEEEcompsocthanksitem Part of this paper's results appeared in IEEE ISIT 2019 \cite{wang2019isit}.}
}

\IEEEtitleabstractindextext{%
\begin{abstract}

Today many mobile users in various zones are invited to sense and send back real-time useful information to keep the freshness of the content updates in such zones. However, due to the sampling cost in sensing and transmission, a user may not have the incentive to contribute real-time information to help reduce the age of information (AoI). We propose dynamic pricing for each zone to offer age-dependent monetary returns and encourage users to sample information at different rates over time. This dynamic pricing design problem needs to well balance the monetary payments as rewards to users and the AoI evolution, and is challenging to solve especially under the incomplete information about users' arrivals and their private sampling costs. After formulating the problem as a nonlinear constrained dynamic program, to avoid the curse of dimensionality, we first propose to approximate the dynamic AoI reduction as a time-average term and successfully solve the approximate dynamic pricing in closed-form. Further, we extend the AoI control from a single zone to many zones with heterogeneous user arrival rates and initial ages, where each zone cares not only its own AoI dynamics but also the average AoI of all the zones in a mean field game system to provide a holistic service. Accordingly, we propose decentralized mean field pricing for each zone to self-operate by using a mean field term to estimate the average age dynamics of all the zones, which does not even require many zones to exchange their local data with each other.

\end{abstract}

}

\maketitle

\IEEEdisplaynontitleabstractindextext

%
\IEEEpeerreviewmaketitle

\section{Introduction}

\IEEEPARstart{C}{ustomers} today prefer not to miss any useful information or breaking news even if in minute, making it imperative for a content provider to keep the posted information fresh for profit \cite{fiveindustries},\cite{google}. The real-time information can be traffic condition, news, sales promotion, and air quality index, which will become outdated and useless over time. To keep information fresh, many content providers now invite and pay the mobile crowd including smartphone users and drivers to sample real-time information frequently \cite{duan2014motivating}. Such crowdsensing approach also saves a content provider's own cost of deploying an expensive sensor network across the city or nation. Recently, the fast development of wireless communication networks and sensors in portable devices enables the mobile users to sample and send back real-time information.

Age of information (AoI) is recently proposed as an important performance metric to quantify the freshness of the information. The literature focuses on the technological issues of the AoI such as the frequency of status updates and queueing delay analysis. In \cite{kaul2012real}, the communication time of the status update systems is considered, and it proves the existence of an optimal packet generation rate at a source to keep its status as timely as possible. The Peak Age-of-Information (PAoI) metric, which is the average maximum age before a new update is received, is considered in \cite{costa2014age} for a single-class M/M/1 queueing system. \cite{sun2017update} shows a counter-intuitive phenomenon that zero-wait policy, i.e., a fresh update is submitted once the previous update is delivered, does not always minimize the age. Considering random packet arrivals, \cite{hsu2017age} and \cite{hsu2017scheduling} study how to keep many customers updated over a wireless broadcast network and a Markov decision process (MDP) is formulated to find dynamic scheduling algorithms. Noting the time-varying availability of energy at the source will affect the update packet transmission rate, \cite{bacinoglu2015age} derives an offline solution that minimizes both the time average age and the peak age for an arbitrary energy replenishment. \cite{zhou2019joint} studies the jointly status sampling and updating processes that can minimize the average AoI at the destination under an average energy cost constraint for each IoT device, and shows that the optimal policy is threshold based related to the AoI state at the device and the AoI state at the destination. \cite{zhou2020age} further discusses the independent device's waiting rate optimization in ultra-dense IoT systems by implementing mean-field games. To analytically prove the optimality of Whittle's index policy for minimizing the AoI, \cite{maatouk2020optimality} introduces a fluid limit model to approximate the behavior of the system and show that the performance of the policy improves as the number of users increases. For a large number of source nodes, \cite{jiang2019unified} applies a mean-field approach to approximate the state distribution over the set of nodes and derives the decentralized status update policy in closed-form. \cite{altman2019forever} considers the users' tradeoff between content update costs and their messages aging, and proposes an age threshold for the users to activate their mobile devices and update information.

However, the economic issues of controlling AoI over different zones are largely overlooked in the literature. On one hand, individuals incur sampling costs when sense and send back their real-time information to content providers, and they should be rewarded and well motivated to contribute their information updates \cite{duan2014motivating}. On the other hand, recruiting a large crowdsensing pool implies a large total sampling cost to compensate, which should be taken into account in a content provider's sustainable management of its AoI (\cite{zhangmeng2019}, \cite{hao2020regulating}). As AoI changes over time, the pricing to reward and motivate users' sensing efforts should be dynamic and age-dependent to optimally balance the AoI evolution and the sampling cost to compensate, yet this new dynamic pricing problem is difficult to solve due to the curse of dimensionality.

Further, we face another challenge for optimally deciding the dynamic pricing in a zone: incomplete information about users' private sampling costs and their random arrivals at the target zone to help sample. Individuals are different in nature and incur different sampling costs to reflect their heterogeneity (e.g., in battery energy consumption and privacy concern when sampling). A user will accept the price offer to sample only if his sampling cost is lower, yet the provider does not know such private cost when deciding and announcing pricing initially. In addition, users' mobilities and their arrivals in different zones to sense are random and different.


Our key novelty and main contributions are summarized as follows.
\begin{itemize}
  \item \emph{Dynamic pricing and mean field analysis for controlling AoI:} To our best knowledge, this paper is the first work studying the dynamic pricing for motivating and controlling AoI update over time, and we take into account the users' random arrivals and their hidden sampling costs. We first look at a single zone to control its individual AoI, by studying how to decide the dynamic pricing to minimize its discounted AoI and monetary payment over time. When extending to the case of many self-operated zones with heterogeneous user arrival rates and initial ages, we further study how to design the decentralized mean field pricing, where each zone cares not only its own AoI dynamics but also the average AoI of all the zones in a mean field game system.
  \item \emph{Approximate dynamic pricing for a single zone:} In the nonlinear constrained dynamic pricing problem of a single zone, we propose to approximate the dynamic AoI reduction as a time-average term and successfully solve the approximate dynamic pricing in closed-form. We prove that if the current AoI is high, a high price offer should be decided to encourage many users to sample. Then we determine the time-average estimator based on Brouwer's fixed-point theorem. We further provide the steady-state analysis for an infinite time horizon, and show that the pricing scheme (though in closed-form) can be further simplified to an $\varepsilon$-optimal version without recursive computing over time.
   \item \emph{Decentralized mean field pricing for multiple zones:} For multiple zones with heterogeneous user arrival rates and initial ages, each zone should further take the average AoI of all the zones into account to provide a holistic service. To reduce the complexity of designing the dynamic pricing for many zones, we propose a mean field term to precisely estimate the average age among all the zones at each time slot. We show that the resulting decentralized dynamic pricing of each zone converges quickly, and a higher user arrival rate results in lower price. When there are a large number of zones, we prove that it is not necessary for zones to exchange their local data with each other, and the system is still ensured stable to reach almost sure asymptotic Nash equilibrium under our mean field pricing.

\end{itemize}

The rest of this paper is organized as follows. The system model and problem formulation are given in Section \ref{sec_systemmodel}. In Section \ref{sec_approximate_dynamic}, we first look at a single zone to control its AoI and solve the approximate dynamic pricing in closed-form. Section \ref{sec_steady_infty} provides the steady-state analysis of our approximate dynamic pricing for an infinite time horizon. The extension to many self-operated zones is studied in Section \ref{sec_Meanfield}. Section \ref{sec_conclusion} concludes this paper.

\subsection{Related Work}

As discussed, the economic issues of controlling AoI in target sensing zones are largely overlooked in the literature. In the few prior works considering the AoI-economics, \cite{zhangmeng2019} discusses how the data providers charge the data users for each update to maximize their profits and how the data users decide their data update schedule to minimize their total costs. \cite{hao2020regulating} studies two platforms' data update schemes under negative network externality due to limited bandwidth. \cite{li2019can} discusses how to incentivize selfish users to provide fresh information by using fixed unit reward. In the other applications of economics and game theory, there are some related works on dynamic pricing (e.g.,\cite{gershkov2014dynamic}--[23]). For example, in \cite{musacchio2006wifi}, dynamic WiFi pricing is discussed under incomplete information of the user's service valuation and utility type. To alleviate data congestion, time-dependent pricing is proposed in \cite{sen2012incentivizing} for Internet service providers to time-shift users' data demand from peak to off-peak periods. \cite{jia2016dynamic} considers the dynamic pricing of electricity in a retail market by finding the tradeoff between consumer surplus and retail profit. In \cite{wang2019dynamic}, the UAV's dynamic service pricing is analyzed by taking into consideration the UAV's energy capacities for hovering and servicing.

However, these works don't consider the system state evolution (e.g., the age dynamics in our problem) affected by the dynamic pricing, not to mention the mean field game framework for large population systems to provide decentralized control. Note that prior works on mean field game systems (e.g., \cite{huang2007large}) and decentralized coordination systems (e.g., \cite{guo2013decentralized}) do not study the dynamic pricing for controlling the time-average state performance. In this paper, we study the dynamic pricing for controlling the weighted sum AoI and analytically solve the decentralized mean field pricing for large zone population.


\begin{figure}
\centering\includegraphics[scale=0.4]{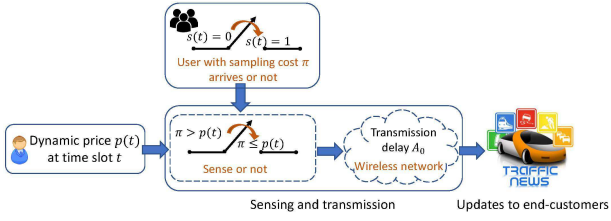}\caption{Illustration of information update process in a particular zone under the zone's dynamic pricing and random user arrival.}\label{fig_systemmodel}
\end{figure}

\section{System Model and Problem Formulation}\label{sec_systemmodel}

We illustrate the information update process for a single zone case in Fig. \ref{fig_systemmodel}. The extension to multiple zones' AoI control in a mean field game will be presented in Sections \ref{sec_Meanfield} and \ref{sec_asympMeanfield}. Here, we consider a discrete time horizon with time slot $t=0,1,\cdots,T$. The provider first announces price $p(t)$ at the beginning of time slot $t$, and a mobile user or sensor may arrive randomly in this target zone in this time slot and (if so) he further decides to sample or not by comparing the price offer $p(t)$ and its own sampling cost $\pi$. If the user appears and accepts to sample ($\pi\leq p(t)$), its sensor data (e.g., about traffic and road condition) is transmitted with random transmission delay $A_0(t)$ to reach the provider's app for end-customers to access. 


We consider that the users (e.g., drivers in a traffic flow) arrive in each zone according to a Poisson process as in the related literature (e.g., \cite{greenshields1946traffic} and \cite{gerlough1955use}) with average number of user arrivals per unit time $\lambda$. Then, the probability of the random number of user arrivals $N(t)$ in the $t$-th time slot of interval $[(t-1)\Delta, t\Delta)$ being equal to $k$ is $Pr(N(t)=k)=\frac{e^{-\lambda\Delta}(\lambda\Delta)^k}{k!}$ with small enough duration $\Delta$ for each time slot. Note that $Pr(N(t)>1)$ becomes trivial as long as we set $\Delta$ small. Thus, each time slot's duration $\Delta$ is properly selected to be short such that it is almost sure to have at most one user arrival at a time.\footnote{For multi-zone case in Section \ref{sec_Meanfield}, to ensure at most one user arrival in a time slot for every zone, the time duration is chosen by $\Delta=\min_{i\in\{1,...,N\}}\Delta_{i}$, where $\Delta_{i}$ is the time duration that at most one user arrive in a time slot for zone $i\in\{1,...,N\}$.} As shown in Fig. \ref{fig_systemmodel}, if a user arrives in time slot $t$, $s(t)=1$; otherwise, $s(t)=0$, with the user arrival rate in each time slot $\alpha=\lambda\Delta$. Further, the users' sampling costs are i.i.d. according to a cumulative distribution function (CDF) $F(\pi), \pi\in[0,b]$, where upperbound $b$ is estimated from historical data. Though all potential users' costs follow the same distribution, their realized costs are different in general. Under the incomplete information, the provider does not know the users' arrivals for potential sampling at time $t$ or the arriving user's particular cost $\pi$. It only knows the user arrival probability $\alpha$ in each time slot and the cost distribution $F(\pi)$. In this paper, we consider non-trivial user arrival rate $\alpha$ and small transmission delay $A_0(t)<1$. Otherwise, the provider should always set the price $p(t)$ to be the upperbound $b$ to encourage the rarely arriving users to sample and reduce the AoI.


\begin{figure}
\centering\includegraphics[scale=0.35]{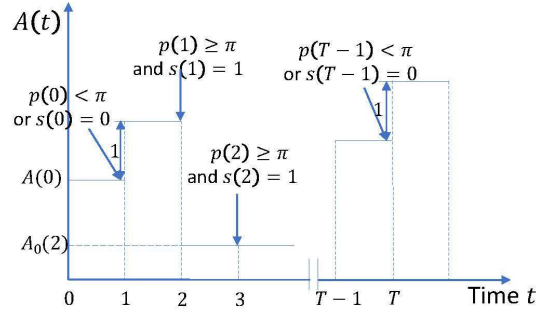}\caption{Actual age $A(t)$ over time under dynamic pricing $p(t)$.}\label{fig_aoimodeltra}
\end{figure}

We adopt the Age of Information (AoI) as the performance metric to quantify the freshness of the information packet at end-customer side. Let $A(t)$ be the instantaneous or realized AoI at the beginning of time slot $t$. Considering a linearly increasing actual age model as in Fig. \ref{fig_aoimodeltra} for the discrete time horizon with age increasement of each time slot normalized to one time slot duration, the new age $A(t+1)$ at time $t+1$ increases from $A(t)$ to $A(t)+1$, if the information is not sampled by any user at time $t$. If a user arrives at time $t$, i.e., $s(t)=1$, and further accepts the price $p(t)$, i.e., $\pi\leq p(t)$, a new status packet will be generated and transmitted. Then, the new age $A(t+1)$ at time $t+1$ drops to $A_0(t)$ by noting the small transmission delay $A_0(t)$ (e.g., in milliseconds) due to small data size of real-time information (e.g., traffic speed). Without loss of generality, we assume the status sampling and transmission are accomplished within a time slot \cite{5guplinkrate}.\footnote{This is reasonable as the information sampling plus transmission delay (in milliseconds) is not comparable to each time slot's duration (in seconds or minutes), and the latter time scale is used to model a new user's physical arrival in the zone.} Then, the dynamics of the instantaneous AoI is given as
\begin{equation}\label{equ_A_r}
A(t+1)=\left\{
\begin{array}{l}
A_0(t), ~~~~~~~\text{if $\pi\leq p(t)$ and $s(t)=1$;} \\
A(t)+1,  ~~\text{otherwise.}\\
\end{array}
\right.
\end{equation}



As long as $A_0(t)$ follows the i.i.d. distribution over any time slot, we denote $A_0=E[A_0(t)]$ as the expected transmission delay of a new data sample. Denote $Pr(A(t+1)|A(t),p(t))$ as the conditional probability to incur AoI $A(t+1)$ at time $t+1$, given price $p(t)$ is applied under instantaneous AoI $A(t)$ at time $t$. According to (\ref{equ_A_r}), given the CDF of an arriving user's cost $F(\pi)$, the conditional probability that the AoI $A(t+1)$ at time $t+1$ will reduce to $A_0(t)$ is $Pr\big(A(t+1)=A_0(t)|A(t),p(t)\big)=\alpha F(p(t))$; and the conditional probability that the AoI $A(t+1)$ at time $t+1$ will increase to $A(t)+1$ is $Pr\big(A(t+1)=A(t)+1|A(t),p(t)\big)=1-\alpha F(p(t))$. Therefore, the expected AoI $E[A(t+1)|A(t),p(t)]$ at time $t+1$ is
\begin{align}\label{equ_AE_dynamic} &E[A(t+1)|A(t),p(t)]\nonumber\\=&E[A_0(t)]Pr(A(t+1)=A_0(t)|A(t),p(t))\nonumber\\
&+(A(t)+1)Pr(A(t+1)=A(t)+1|A(t),p(t))\nonumber\\
=&A_0\alpha F(p(t))+(A(t)+1)(1-\alpha F(p(t))). \end{align}

For ease of reading, we rewrite the conditional expectation in (\ref{equ_AE_dynamic}) as $E[A(t+1)]$. Considering uniform distribution for the users' private costs (i.e., $F(\pi)=\frac{\pi}{b}$, $\pi\in [0,b]$), according to (\ref{equ_AE_dynamic}), the dynamics of the expected age to next time is given as:
\begin{align}\label{equ_A_dynamic} E[A(t+1)]=&A_0\alpha \frac{p(t)}{b}+(E[A(t)]+1)(1-\alpha \frac{p(t)}{b})\nonumber\\
=&E[A(t)]-(E[A(t)]-A_0)\alpha \frac{p(t)}{b}\nonumber\\
&+(1-\alpha\frac{p(t)}{b}), \end{align}
which shows nonlinear AoI evolution due to the product of the expected age $E[A(t)]$ and $p(t)$ in the second term above.


Since the probability that a user appears and accepts the price $p(t)$ is $\frac{\alpha p(t)}{b}$, the expected payment to this user is $\frac{\alpha p^2(t)}{b}$. Note that the optimal price $p(t)$ should not exceed the maximum cost $b$ of the user as it is unnecessary for the provider to over-pay. The objective of the provider is to determine the optimal dynamic pricing $p(t), t\in\{0,...,T\}$ that minimizes the expected total discounted cost over time, which is the summation of the square expected age and expected monetary payment over time:
\bee\label{equ_UTobjective} U(T)=\min_{p(t)\in [0,b],t\in\{0,...,T\}}\sum_{t=0}^{T}\rho^t(E[A(t)]^2+\frac{\alpha p^2(t)}{b}), \ene

\begin{align} \text{s.t.}~~E[A(t+1)]=&E[A(t)]-(E[A(t)]-A_0)\alpha \frac{p(t)}{b}\nonumber\\
&+(1-\alpha\frac{p(t)}{b})\nonumber, ~~~~~~~~~~~~~~~~~~~~~~~~~~~~~~~~~~~\text{(\ref{equ_A_dynamic})}\end{align}

\noindent where $\rho\in(0,1)$ is the discount factor and indicates the provider values the current costs than that of future. We choose the typical square expected age $E[A(t)]^2$ as in many economics papers \cite{microEcoTheory} to reflect the fact that the platform provider's profit loss should be convexly increasing in the age of its provided information.\footnote{The end-customers are sensitive to information freshness and many media platforms (e.g., Netflix, navigation apps) want to keep provided information as fresh as possible. If the AoI is large, the user experience is bad due to outdated information and the end-customers may switch to other providers, which leads to profit loss.}


The problem in (\ref{equ_A_dynamic})-(\ref{equ_UTobjective}) is a nonlinear constrained dynamic program, which is challenging to solve analytically due to the curse of dimensionality. Imagining that a huge number of time-dependent prices $p(t)\in[0,b], t\in\{0,...,T\}$ must be jointly designed with the price interval $\tau$, then the computation complexity $O((\frac{b}{\tau})^T)$ is formidably high and increases exponentially in $T$. To analytically solve this problem for useful insights and guidelines to the provider, we will propose an approximation of dynamic pricing in the following section.

\section{Approximation of Dynamic Pricing}\label{sec_approximate_dynamic}

To analytically obtain the optimal dynamic prices $p(t), t\in\{0,...,T\}$ for the nonlinear constrained dynamic problem in (\ref{equ_A_dynamic})-(\ref{equ_UTobjective}), we approximate the nonlinear dynamics constraint of the expected age in (\ref{equ_A_dynamic}) into linear dynamics. Specifically, we propose a time-average term $\delta$ as an estimator to approximate the possible age reduction $E[A(t)]-A_0$ at each time slot $t$ in (\ref{equ_A_dynamic}), i.e.,
\bee\label{equ_A_dynamic_estimate} E[A(t+1)]=E[A(t)]-\delta\alpha \frac{p(t)}{b}+(1-\alpha \frac{p(t)}{b}). \ene
Here, the time-average estimator $\delta$ is viewed as the time-average age reduction of each time slot and its estimation should also take into account the time discount factor $\rho$, i.e.,\footnote{There is no need to include $E[A(T)]$ in $\delta$ estimation, as it will not affect the AoI $E[A(t)]$ in any previous time slot $t\leq T-1$.}
\bee\label{equ_delta} \delta=\frac{1-\rho}{1-\rho^T}\sum_{t=0}^{T-1}\rho^t (E[A(t)]-A_0). \ene
Note that when time discount factor $\rho\rightarrow 1$, by applying L'Hôpital's rule, we have $\delta=\lim_{\rho\rightarrow 1}\frac{1-\rho}{1-\rho^T}\sum_{t=0}^{T-1}\rho^t (E[A(t)]-A_0)=\frac{1}{T}\sum_{t=0}^{T-1}(E[A(t)]-A_0)$.


In the following, given any estimator $\delta$, we analyze the approximate dynamic pricing in Section \ref{sec_approx_pricing1}. Later in Section \ref{sec_fixed_delta}, we will further analyze how to determine the estimator $\delta$ for finalizing pricing update. We will also show that this approximate dynamic pricing approaches the optimal dynamic pricing well.

\subsection{Analysis of Approximate Dynamic Pricing}\label{sec_approx_pricing1}

Though the dynamic programming problem (\ref{equ_UTobjective})-(\ref{equ_A_dynamic_estimate}) now has only linear AoI evolution constraint in (\ref{equ_A_dynamic_estimate}), it is still not easy to solve by considering the huge number of price combinations over time. We denote the cost objective function from initial time $t$ as
\bee\label{equ_J_from_t} \begin{split} J(p,t)
=\sum_{s=t}^{T}\rho^{s-t}(E[A(s)]^2+\frac{\alpha}{b}p^2(s)),
\end{split}\ene
and the value function given the initial age $A(t)$ as
\bee\label{equ_Vi} V(E[A(t)],t)=\min_{\{p(s)\in [0,b]\}_{s=t}^T}(J(p,t)|E[A(t)]). \ene
Then, we have the dynamic programming equation at time $t$:
\bee\label{equ_VAt}\begin{split} V(E[A(t)],t)=&\min_{p(t)\in [0,b]}E[A(t)]^2+\frac{\alpha}{b}p^2(t)\\&+\rho V(E[A(t+1)],t+1),\end{split}\ene
\rightline{\text{s.t.}~~$E[A(t+1)]=E[A(t)]-\delta\alpha \frac{p(t)}{b}+(1-\alpha\frac{p(t)}{b}).$ ~~~~(\ref{equ_A_dynamic_estimate})}

\begin{figure}[htbp]
\centering
\subfigure[Approximate dynamic pricing $p(t)$ versus time $t$.]{\label{subfig_p(t)low}
\includegraphics[height=1.9in,width=2.4in,angle=0]{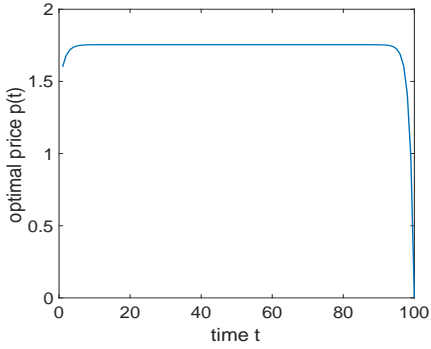}
}\\
\subfigure[Expected age \text{$E[A(t)]$} versus time $t$.]{\label{subfig_agelow}
\includegraphics[height=1.9in,width=2.4in,angle=0]{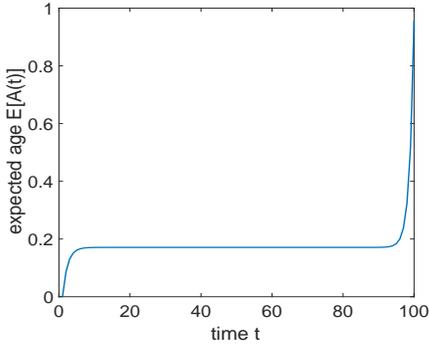}
}
\caption{Dynamics of price approximation $p(t)$ and expected age $E[A(t)]$ over $T=100$ time slots when $A(0)=0, \alpha=1, b=2, \rho=0.9, \delta=0.14$ (obtained according to Algorithm \ref{alg_find_fixed_b1}).}\label{fig_dynamics}
\end{figure}

In the following, we will analyze the optimal solution to the dynamic problem above by using dynamic control techniques.

\begin{pro}\label{pro_optimal_pt} The approximate dynamic pricing $p(t), t\in\{0,...,T\}$ as the optimal solution to the dynamic program (\ref{equ_UTobjective})-(\ref{equ_A_dynamic_estimate}) is monotonically increasing in $E[A(t)]$ and given by\footnote{Note that (\ref{equ_pt_finite}) automatically satisfies the constraint $p(t)\in [0, b]$ as long as the transmission delay $A_0$ is not huge and the zone has non-trivial user arrival rate $\alpha$. In extreme case of $p(t)>b$ for the first few time slots (e.g., due to huge initial age $A(0)$), we can simply reset $p(t)$ to be the upperbound $b$ and expected age $A(t)$ will decrease and ensure $p(t)\leq b$ in the long run.}
\bee\label{equ_pt_finite} p(t)=\frac{\rho M_{t+1}(\delta+1)+2\rho(\delta+1)Q_{t+1}(E[A(t)]+1)}{2+2\rho Q_{t+1}\frac{\alpha(\delta+1)^2}{b}}, \ene
with $p(T)=0$, and the resulting expected age $E[A(t)]$ at time $t$ is
\bee\label{equ_At_non_infty}\begin{split} E[A(t)]=&\prod_{i=1}^t\frac{1}{1+\rho Q_i\frac{\alpha(\delta+1)^2}{b}}A(0)+\frac{2-\rho M_t\frac{\alpha(\delta+1)^2}{b}}{2+2\rho Q_t\frac{\alpha(\delta+1)^2}{b}}\\
+&\sum_{s=1}^{t-1}\frac{2-\rho M_s\frac{\alpha(\delta+1)^2}{b}}{2+2\rho Q_s\frac{\alpha(\delta+1)^2}{b}}\prod_{i=s+1}^t\frac{1}{1+\rho Q_i\frac{\alpha(\delta+1)^2}{b}}, \end{split}\ene
where
\bee\label{equ_Qt} Q_t=1+\frac{\rho Q_{t+1}}{1+\rho Q_{t+1}\frac{\alpha(\delta+1)^2}{b}}, \ene
\bee\label{equ_Mt} M_t=\frac{\rho(M_{t+1}+2 Q_{t+1})}{1+\rho Q_{t+1}\frac{\alpha(\delta+1)^2}{b}}, \ene
with $Q_T=1, M_T=0$ on the boundary.
\end{pro}

\textbf{Proof Sketch:}  According to the first-order condition $\frac{\partial V(E[A(t)],t)}{\partial p(t)}=0$ when solving (\ref{equ_VAt}) in the backward induction process, we notice that $p(t)$ is a linear function of $E[A(t)]$. Thus, the value function in (\ref{equ_VAt}) should follow the following quadratic structure with respect to $E[A(t)]$:
\bee\label{equ_VAguess} V(E[A(t)],t)=Q_tE[A(t)]^2+M_tE[A(t)]+S_t, \ene
yet we still need to determine $Q_t, M_t, S_t$. This will be accomplished by finding the recursion in the following.

First, we have $Q_T=1, M_T=0, S_T=0$ on the boundary due to $V(E[A(T)],T)=E[A(T)]^2$. Given $V(E[A(t+1)],t+1)=Q_{t+1}E[A(t+1)]^2+M_{t+1}E[A(t+1)]+S_{t+1}$ as in (\ref{equ_VAguess}), the dynamic programming equation at time $t$ is
\bee\label{equ_Vt}\begin{split} &V(E[A(t)],t)\\=&\min_{p(t)}\Big(E[A(t)]^2+\frac{\alpha}{b}p^2(t)+\rho Q_{t+1}E[A(t+1)]^2\\&+\rho M_{t+1}E[A(t+1)]+\rho S_{t+1}\Big).\end{split}\ene

Substitute $E[A(t+1)]$ in (\ref{equ_A_dynamic_estimate}) into (\ref{equ_Vt}) and let $\frac{\partial V(E[A(t)],t)}{\partial p(t)}=0$, we obtain the optimal price $p(t)$ in (\ref{equ_pt_finite}). Then, we substitute $p(t)$ in (\ref{equ_pt_finite}) into $V(E[A(t)],t)$ in (\ref{equ_Vt}), and obtain $V(E[A(t)],t)$ as a function of $Q_{t+1}, M_{t+1}, S_{t+1}$ and $E[A(t)]$. Finally, by reformulating $V(E[A(t)],t)$ in (\ref{equ_Vt}) and noting that $V(E[A(t)],t)=Q_tE[A(t)]^2+M_tE[A(t)]+S_t$, we obtain the recursive functions of $Q_t$ and $M_t$ in (\ref{equ_Qt}) and (\ref{equ_Mt}). Substitute $p(t)$ in (\ref{equ_pt_finite}) into (\ref{equ_A_dynamic_estimate}), we obtain the expected age $E[A(t)]$ in (\ref{equ_At_non_infty}). \qed


Note that $p(t)$ in (\ref{equ_pt_finite}) also applies to the case without discount factor (i.e., $\rho=1$) and we expect a less myopic pricing to control future age increment. Fig. \ref{fig_dynamics} simulates the system dynamics over time under the approximate dynamic price $p(t)$ in (\ref{equ_pt_finite}). We can see that when the initial age $A(0)$ is low, $p(t)$ is low initially to save sampling cost and can allow the age to increase mildly. The price $p(t)$ then increases with the increased age until both of them reach steady-states. This is consistent with the monotonically increasing relationship between $p(t)$ and $E[A(t)]$ in (\ref{equ_pt_finite}). Yet when close to the end of the time horizon $T=100$, the price $p(t)$ decreases to $0$ to save immediate sampling expense without worrying its effect to increase the future age. Consequently, the expected age $E[A(t)]$ increases again but only lasts for a few time slots.

\subsection{Update of Estimator $\delta$ for Finalizing Pricing}\label{sec_fixed_delta}


Now we are ready to update the estimator $\delta$ defined in (\ref{equ_delta}) for finalizing dynamic pricing in (\ref{equ_pt_finite}). The estimator $\delta=\frac{1-\rho}{1-\rho^T}\sum_{t=0}^{T-1}\rho^t (E[A(t)]-A_0)$ is affected by all the estimated ages $E[A(t)]$ over the time horizon $t\in\{0,...,T-1\}$, which will in turn affect $E[A(t)]$ in (\ref{equ_At_non_infty}) in the closed loop. Thus we next determine $\delta$ by finding the fixed point of (\ref{equ_delta}), which will be shown to exist in our problem.

\begin{algorithm}[t]
\caption{Iterative computation of fixed point estimator $\delta$ in (\ref{equ_delta})}
\begin{algorithmic}[1]

\STATE Initiate $\epsilon'=1,j=1,\epsilon=0.001$, an arbitary initial $\delta^{est}(0)\geq 0$, and $\delta=\delta^{est}(0)$

\WHILE {$\epsilon'>\epsilon$}

\FOR {$t=0$ to $T-1$}
\STATE Compute $Q_t$ and $M_t$ according to $\delta$, (\ref{equ_Qt}), (\ref{equ_Mt})
\ENDFOR

\FOR {$t=1$ to $T-1$}
\STATE Compute $E[A(t)]$ according to (\ref{equ_At_non_infty})
\ENDFOR

\STATE $\delta^{est}(j)=\frac{1-\rho}{1-\rho^T}\sum_{t=0}^{T-1}\rho^t (E[A(t)]-A_0)$
\STATE $\delta=\delta^{est}(j)$
\STATE $\epsilon'=\delta^{est}(j)-\delta^{est}(j-1)$
\STATE $j=j+1$

\ENDWHILE
\RETURN Fixed point $\delta$

\end{algorithmic}
\label{alg_find_fixed_b1}
\end{algorithm}

\begin{figure}
\centering\includegraphics[scale=0.285]{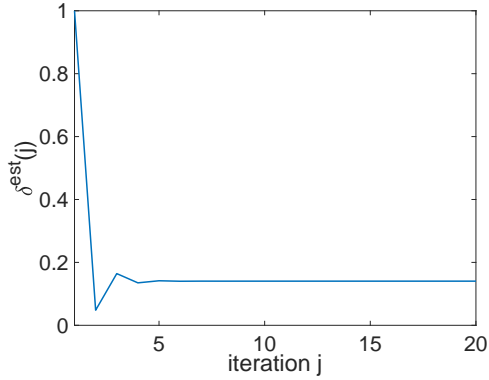}\caption{Convergence of average age estimator $\delta^{est}(j)$ to the fixed point $\delta$ by Algorithm \ref{alg_find_fixed_b1} when $T=100$, $A(0)=0, \alpha=1, b=2, \rho=0.9$.}\label{fig_aoideltaConverge}
\end{figure}

For any $1\leq t \leq T-1$, we substitute $\delta=\frac{1-\rho}{1-\rho^T}\sum_{t=0}^{T-1}\rho^t (E[A(t)]-A_0)$ in (\ref{equ_delta}) into $E[A(t)]$ in (\ref{equ_At_non_infty}), and define
\bee\label{equ_phitA}\begin{split} E[A(t)]=\Phi_t(E[A(1)],\cdots,E[A(T-1)]),  \end{split}\ene
which tells $E[A(t)]$ is a function of all the ages $\{E[A(t)], t\in\{1, ..., T-1\}\}$.
Then, we summarize by using the following vector function $\Phi$ for $t\in\{1,\cdots,T-1\}$:
\bee\label{equ_Phivec}\begin{split} &\Phi(E[A(1)],\cdots,E[A(T-1)])\\=&(\Phi_1(E[A(1)],\cdots,E[A(T-1)]),\cdots,\\&\Phi_{T-1}(E[A(1)],\cdots,E[A(T-1)])). \end{split}\ene
The fixed point (if any) in $\Phi(E[A(1)],\cdots,E[A(T-1)])=(E[A(1)],\cdots,E[A(T-1)])$ should be reached to let $\delta$ replicate $\frac{1-\rho}{1-\rho^T}\sum_{t=0}^{T-1}\rho^t (E[A(t)]-A_0)$ in (\ref{equ_delta}) in the first place.


Note that $Q_t\geq 1$ in (\ref{equ_Qt}) and $M_t\geq 0$ in (\ref{equ_Mt}), we have the following upperbound for (\ref{equ_phitA}):
\bee \Phi_t\leq A(0)+t. \ene Define continuous space  $\Omega=[0,A(0)+1]\times\cdots\times[0,A(0)+T-1]$ in $T-1$ dimensions. Since each $\Phi_t$ in (\ref{equ_phitA}) is continuous in $\Omega$, the vector function $\Phi$ in (\ref{equ_Phivec}) is a continuous mapping from $\Omega$ to $\Omega$. According to the Brouwer's fixed-point theorem, we have the following proposition.


\begin{pro} $\Phi(E[A(1)],\cdots,E[A(T-1)])$ in (\ref{equ_Phivec}) has a fixed point in $\Omega$.
\end{pro}

Given the existence of the fixed point, we are ready to find the estimator $\delta$ in (\ref{equ_delta}). Accordingly, we propose Algorithm \ref{alg_find_fixed_b1}: given any initial estimator $\delta^{est}(j)$ in round $j$, we can iteratively obtain the resulting expected ages $E[A(t)], t\in\{1,\cdots,T-1\}$ according to (\ref{equ_At_non_infty}), and then check whether the resulting estimator $\delta^{est}(j+1)$ in next round converges to the last-round estimator $\delta^{est}(j)$. By repeating the iteration process until $\delta^{est}(j+1)$ is in the neighborhood of $\delta^{est}(j)$ with error $\epsilon$, we obtain the fixed point $\delta$ and the computation complexity of Algorithm \ref{alg_find_fixed_b1} is $O(\frac{T}{\epsilon})$. As shown in Fig. \ref{fig_aoideltaConverge}, $\delta^{est}(j)$ converges to the fixed point $\delta=0.14$ within $7$ iterations. After obtaining the value of $\delta$ along with $Q_t$ and $M_t, t\in\{0,\cdots,T\}$, we just need to substitute them into the closed-form solution in (\ref{equ_pt_finite}) and obtain approximate dynamic pricing. 

\begin{figure}
\centering\includegraphics[scale=0.29]{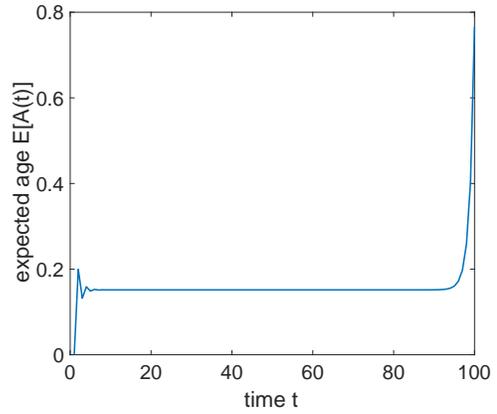}\caption{Convergence of expected age $E[A(t)]$ in the original nonlinear AoI dynamics system in (\ref{equ_A_dynamic}) under approximate dynamic pricing $p(t)$ in (\ref{equ_pt_finite}). The values of the parameters are set the same as in Fig. \ref{subfig_agelow}.}\label{fig_staticAoI}
\end{figure}

\begin{figure}
\centering\includegraphics[scale=0.27]{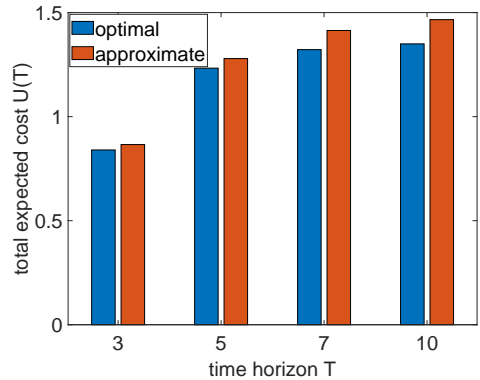}\caption{The gap between the total expected costs under the optimal pricing (obtained via exhaustive search) and our approximate pricing (obtained in (\ref{equ_pt_finite}) and Algorithm \ref{alg_find_fixed_b1}) in the original system (\ref{equ_A_dynamic})-(\ref{equ_UTobjective}) when $A(0)=0, \alpha=1, b=2, \rho=0.5$.}\label{fig_costgap}
\end{figure}


In the following, we examine the performance of the approximate dynamic pricing $p(t)$ in (\ref{equ_pt_finite}) in the original AoI dynamics system (\ref{equ_A_dynamic}) without linearizing the age reduction in (\ref{equ_A_dynamic_estimate}). We wonder if the approximate pricing obtained after linearization still ensures the expected age $E[A(t)]$ in the original AoI dynamics to converge and reach a steady-state. If so, we want to further examine the performance gap between the total expected costs under the optimal dynamic price (obtained via exhaustive search) and our approximate dynamic price in the original system.

First, by applying the approximate dynamic pricing $p(t)$ in (\ref{equ_pt_finite}) to the original AoI dynamics in (\ref{equ_A_dynamic}), Fig. \ref{fig_staticAoI} shows that the expected age $E[A(t)]$ still converges to a steady-state only after a few time slots, which shows that our approximate pricing is feasible to control the AoI evolution in the original system. This figure is similar to Fig. \ref{subfig_agelow} in the linearized AoI dynamics system (under the same parameter setting). Thus, we can tell that the time-average estimator $\delta$ approximates the age evolution well.

Although we may imagine that the approximate dynamic pricing will cause performance losses (though saving computational complexity greatly), Fig.~\ref{fig_costgap} shows that the total expected cost gap between the optimal dynamic pricing and our approximate dynamic pricing in the original system (\ref{equ_A_dynamic})-(\ref{equ_UTobjective}) is small. 

\subsection{Extension to Other Distributions of Users' Private Costs}\label{sec_normalDistribution}


\begin{figure}
\centering\includegraphics[scale=0.24]{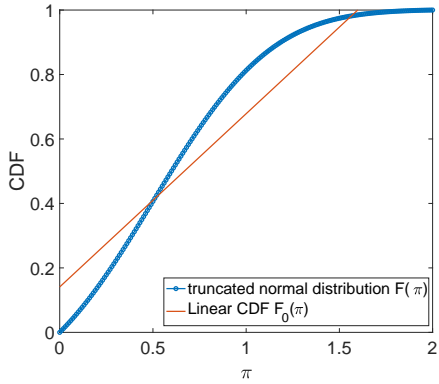}\caption{Approximation of truncated normal distribution $F(\pi)$ with $F_0(\pi)=0.14+0.54\pi$ when $\mu=0.5, \sigma=2, b=2$ with norm of residuals $1.46$.}\label{fig_LinearNormalDistribution}
\end{figure}

Our results can also be extended to some other continuous distributions of users' private costs with CDF function $F(\pi)$ by applying the first-order approximation on $F(\pi)$. To be specific, we first use a linear function $F_0(\pi)=a_1+a_2\pi$ to approximate $F(\pi)$, where $a_1$ and $a_2$ are coefficients returned by linear regression. Take the truncated normal distribution for example, it has original CDF:
\bee\label{equ_normaldistribution} F(\pi)=\frac{erf(\frac{\pi-\mu}{\sqrt{2}\sigma})-erf(\frac{-\mu}{\sqrt{2}\sigma})}{erf(\frac{b-\mu}{\sqrt{2}\sigma})-erf(\frac{-\mu}{\sqrt{2}\sigma})}, \pi\in [0,b],\ene
with mean $\mu$ and variance $\sigma^2$. As shown in Fig. \ref{fig_LinearNormalDistribution}, we approximate it as the linear function $F_0(\pi)=0.14+0.54\pi$, which is truncated within $[0,1]$.

\begin{figure}
\centering\includegraphics[scale=0.265]{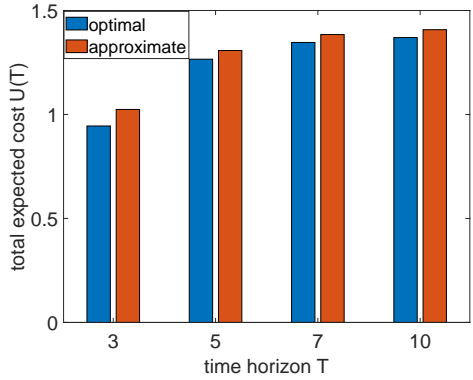}\caption{The gap between the total expected costs under the optimal pricing (obtained via exhaustive search with truncated normal distribution $F(\pi)$ in (\ref{equ_normaldistribution})) and our approximate dynamic pricing in (\ref{equ_pt_finite}) (with the first-order approximation of the CDF $F_0(\pi)$) in the original nonlinear system.}\label{fig_GapNormalDistribution}
\end{figure}

Then, based on the linear approximation $F_0(\pi)$, similar to our analysis in Sections \ref{sec_approx_pricing1} and \ref{sec_fixed_delta}, we can obtain the approximate dynamic pricing $p(t)$ as in (\ref{equ_pt_finite}). 
It is shown in Fig. \ref{fig_GapNormalDistribution} that, for the truncated normal distribution of users' private costs, the total expected cost gap between the optimal dynamic pricing (obtained via exhaustive search with original CDF $F(\pi)$ in (\ref{equ_normaldistribution})) and our approximate dynamic pricing in (\ref{equ_pt_finite}) (with the first-order approximation of the CDF $F_0(\pi)$) in the original nonlinear system is small.


\section{Steady-state Analysis of Dynamic Pricing}\label{sec_steady_infty}



\begin{figure}
\centering\includegraphics[scale=0.28]{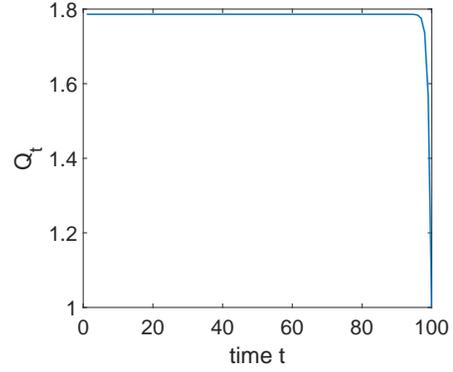}\caption{$Q_t$ in (\ref{equ_Qt}) versus time $t$ with $T=100$ time slots.}\label{fig_Q_t}
\end{figure}

\begin{figure}
\centering\includegraphics[scale=0.28]{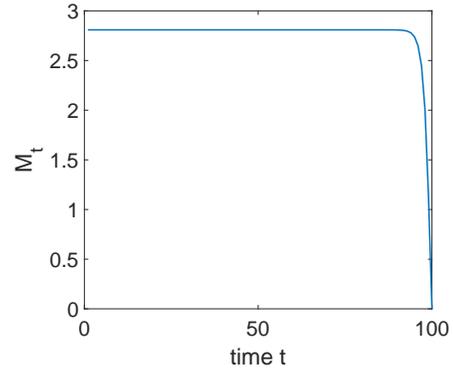}\caption{$M_t$ in (\ref{equ_Mt}) versus time $t$ with $T=100$ time slots.}\label{fig_M_t}
\end{figure}

We wonder how our approximate dynamic pricing and its performance would be in the steady-state, by looking at the infinite time horizon $T\rightarrow\infty$ in this section. As $T\rightarrow \infty$, there is no ending time and we do not observe the surge of pricing $p(t)$ and age $E[A(t)]$ as in Figs. \ref{fig_dynamics} and \ref{fig_staticAoI} when closing to the end $T$. We will show that the approximate dynamic pricing can be further simplified to an $\varepsilon$-optimal version without recursive computing in (\ref{equ_pt_finite}) over time. Specifically, the steady-state characterizations of $Q_t$ in (\ref{equ_Qt}) and $M_t$ in (\ref{equ_Mt}) can be found by iterating the dynamic equations until they converge. The following lemma shows the steady-states of $Q_t$ and $M_t$, both of which exist and are nicely given in closed-form.

\begin{lem}\label{lem_Mt_Q_t_infinity} As $T\rightarrow\infty$, $Q_t$ in (\ref{equ_Qt}) and $M_t$ in (\ref{equ_Mt}) respectively converge to the following steady-states: \begin{align}\label{equ_solve_Q} Q=&\frac{1}{2}\Big(1-\frac{b(1-\rho)}{\rho \alpha(\delta+1)^2}\nonumber\\
&+\sqrt{(1-\frac{b(1-\rho)}{\rho \alpha(\delta+1)^2})^2+\frac{4b}{\rho\alpha(\delta+1)^2}}\Big), \\
 M=&\frac{2\rho Q}{1-\rho+\rho Q\frac{\alpha(\delta+1)^2}{b}}. \label{equ_stableM}\end{align}
\end{lem}

\textbf{Proof:} Starting from the boundary condition $Q_T=1, M_T=0$, according to $Q_{t}=1+\frac{\rho Q_{t+1}}{1+\rho Q_{t+1}\frac{\alpha(\delta+1)^2}{b}}$ in (\ref{equ_Qt}) and $M_t=\frac{\rho(M_{t+1}+2 Q_{t+1})}{1+\rho Q_{t+1}\frac{\alpha(\delta+1)^2}{b}}$ in (\ref{equ_Mt}), we can obtain $Q_t$ and $M_t$ backward given $Q_{t+1}$ and $M_{t+1}$ at next time slot. Rewrite $Q_{t}$ above as
\bee\label{equ_QtEvolve} Q_{t}=1+\frac{\rho}{\frac{1}{Q_{t+1}}+\rho\frac{\alpha(\delta+1)^2}{b}}. \ene
Starting from $Q_T=1$ and note that $\frac{\rho}{\frac{1}{Q_{t+1}}+\rho\frac{\alpha(\delta+1)^2}{b}}>0$, we have $Q_{T-1}>1>Q_T$. Note that (\ref{equ_QtEvolve}) is increasing with $Q_{t+1}$ and $\lim_{Q_{t+1}\rightarrow\infty}=1+\frac{b}{\alpha(\delta+1)^2}<+\infty$, we can conclude that $\{Q_T, Q_{T-1}, ..., Q_0\}$ is bounded increasing sequence in the reverse time order and will converge to the steady-state $Q$, which can be solved as (\ref{equ_solve_Q}) by removing the time subscripts from (\ref{equ_Qt}).

\begin{figure}
\centering\includegraphics[scale=0.3]{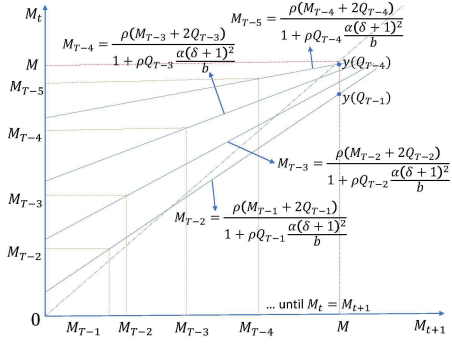}\caption{Evolution of $M_t$.}\label{fig_Q_tConvergeProof}
\end{figure}

By removing the time subscripts from (\ref{equ_Mt}), the steady-state $M$ is solved as $M=\frac{2\rho Q}{1-\rho+\rho Q\frac{\alpha(\delta+1)^2}{b}}$ in (\ref{equ_stableM}). In the following, we will show that $\{M_T, M_{T-1}, ..., M_0\}$ is bounded increasing sequence in the reverse time order and converges to $M$. According to $M_t=\frac{\rho(M_{t+1}+2 Q_{t+1})}{1+\rho Q_{t+1}\frac{\alpha(\delta+1)^2}{b}}$ in (\ref{equ_Mt}), define $y(Q_{t+1})$ as the right-hand side of (\ref{equ_Mt}) given the steady-state $M$, i.e.,
\bee y(Q_{t+1})=\frac{\rho(M+2 Q_{t+1})}{1+\rho Q_{t+1}\frac{\alpha(\delta+1)^2}{b}}. \ene

Take the derivative of $y(Q_{t+1})$ with respect to $Q_{t+1}$, we have
\bee \frac{dy(Q_{t+1})}{dQ_{t+1}}=\frac{2\rho-M\rho^2\frac{\alpha(\delta+1)^2}{b}}{(1+\rho Q_{t+1}\frac{\alpha(\delta+1)^2}{b})^2}. \ene
According to $M=\frac{2\rho Q}{1-\rho+\rho Q\frac{\alpha(\delta+1)^2}{b}}$, we have $M\rho\frac{\alpha(\delta+1)^2}{b}=\frac{2\rho Q\rho\frac{\alpha(\delta+1)^2}{b}}{1-\rho+\rho Q\frac{\alpha(\delta+1)^2}{b}}=\frac{2\rho}{\frac{1-\rho}{Q\rho\frac{\alpha(\delta+1)^2}{b}}+1}<2$. Thus, $\frac{dy(Q_{t+1})}{dQ_{t+1}}>0$ and $y(Q_{t+1})$ increases with $Q_{t+1}$. Since $\{Q_T, Q_{T-1}, ..., Q_0\}$ is an increasing sequence and converges to $Q$, $y(Q_{T})<y(Q_{T-1})<\cdots<y(Q)$.

Starting from $M_T=0$, we have $M_{T-1}=\frac{2\rho}{1+\rho\frac{\alpha(\delta+1)^2}{b}}>0=M_T$. Since $\frac{2\rho Q_{t+1}}{1+\rho Q_{t+1}\frac{\alpha(\delta+1)^2}{b}}$ increases with $Q_{t+1}$, as shown in Fig. \ref{fig_Q_tConvergeProof}, the equation set $\{M_t=\frac{\rho(M_{t+1}+2 Q_{t+1})}{1+\rho Q_{t+1}\frac{\alpha(\delta+1)^2}{b}}\}$ intersects with the vertical axis with $\frac{2\rho Q_{T-1}}{1+\rho Q_{T-1}\frac{\alpha(\delta+1)^2}{b}}\leq\frac{2\rho Q_{T-2}}{1+\rho Q_{T-2}\frac{\alpha(\delta+1)^2}{b}}\leq\cdots\leq\frac{2\rho Q_{T-4}}{1+\rho Q_{T-4}\frac{\alpha(\delta+1)^2}{b}}$. The slope of each equation satisfies $0<\frac{\rho}{1+\rho Q_{t+1}\frac{\alpha(\delta+1)^2}{b}}<1$. Therefore, starting from $M_{T-1}>0$, we have $M_{T-2}>M_{T-1}$ based on $M_{T-2}=\frac{\rho(M_{T-1}+2 Q_{T-1})}{1+\rho Q_{T-1}\frac{\alpha(\delta+1)^2}{b}}$. Then, from $M_{T-2}$, we have $M_{T-3}>M_{T-2}$ based on $M_{T-3}=\frac{\rho(M_{T-3}+2 Q_{T-3})}{1+\rho Q_{T-3}\frac{\alpha(\delta+1)^2}{b}}$. Thus, we can obtain $M_{T}<M_{T-1}<M_{T-2}<M_{T-3}<\cdots$ until $M_t=M_{t+1}=M$. \qed

Figs. \ref{fig_Q_t} and \ref{fig_M_t} simulate the dynamics of $Q_t$ in (\ref{equ_Qt}) and $M_t$ in (\ref{equ_Mt}). We can see that both $Q_t$ and $M_t$ fast converge to their steady-states in a few rounds.

According to $Q$ in (\ref{equ_solve_Q}) and $M$ in (\ref{equ_stableM}), and note that $\frac{1}{1+\rho Q\frac{\alpha(\delta+1)^2}{b}}<1$, we have the following proposition.

\begin{pro}\label{pro_steadyprice} For the infinite time horizon case, the approximate pricing is simplified from (\ref{equ_pt_finite}) to
\bee\label{equ_pt_infinity_horizon} p^{\infty}(t)=\frac{\rho M(\delta+1)+2\rho(\delta+1)Q(E[A^{\infty}(t)]+1)}{2+2\rho Q\frac{\alpha(\delta+1)^2}{b}}, \ene

\noindent and the resulting expected age $A^{\infty}(t)$ at time $t$ is
\begin{align}\label{equ_At_forinfty} E[A^{\infty}(t)]=&(\frac{1}{1+\rho Q\frac{\alpha(\delta+1)^2}{b}})^tA(0)\nonumber\\
&+\frac{2-\rho M\frac{\alpha(\delta+1)^2}{b}}{2+2\rho Q\frac{\alpha(\delta+1)^2}{b}}\frac{1-(\frac{1}{1+\rho Q\frac{\alpha(\delta+1)^2}{b}})^t}{1-\frac{1}{1+\rho Q\frac{\alpha(\delta+1)^2}{b}}}. \end{align}

As $t\rightarrow\infty$, the expected age converge to
\bee\label{equ_At_infty}\begin{split} \lim_{t\rightarrow\infty}E[A^{\infty}(t)]=\frac{(1-\rho)(1+\rho Q\frac{\alpha(\delta+1)^2}{b})}{\rho Q(\delta+1)^2(\frac{\alpha}{b}(1-\rho)+\rho Q(\frac{\alpha(\delta+1)}{b})^2)}, \end{split}\ene
and the optimal dynamic pricing converges to
\bee\label{equ_pt_infty} \lim_{t\rightarrow\infty}p^{\infty}(t)=\frac{b}{\alpha(\delta+1)}. \ene
\end{pro}



In the following, we prove the existence and uniqueness of the fixed point estimator $\delta$ for the infinite time horizon. 

\begin{pro}\label{pro_fixedpoint} For the infinite time horizon, the estimator $\delta$ exists and can be solved as the unique positive solution to the following equation:
\bee\label{equ_solve_b1rewrite} \frac{\rho Q\frac{\alpha(\delta+1)^2}{b}(\delta+A_0)(1-\rho+\rho Q\frac{\alpha(\delta+1)^2}{b})}{(1-\rho)(1+\rho Q\frac{\alpha(\delta+1)^2}{b})}=1, \ene
where $Q$ in (\ref{equ_solve_Q}) is a function of $\delta$.
\end{pro}

\textbf{Proof:} Note that $\delta$ is estimated by $\frac{1-\rho}{1-\rho^T}\sum_{t=0}^{T-1}\rho^t (E[A^{\infty}(t)]-A_0)$ in (\ref{equ_delta}). Since $E[A^{\infty}(t)]$ converges to (\ref{equ_At_infty}), as $T\rightarrow\infty$, we can further estimate $\delta$ as
\begin{align}\label{equ_deltasolve} \delta=&\frac{1-\rho}{1-\rho^T}\sum_{t=0}^{T-1}\rho^t (E[A^{\infty}(\infty)]-A_0)\nonumber\\
=&\frac{(1-\rho)(1+\rho Q\frac{\alpha(\delta+1)^2}{b})}{\rho Q(\delta+1)^2(\frac{\alpha}{b}(1-\rho)+\rho Q(\frac{\alpha(\delta+1)}{b})^2)}-A_0. \end{align}

Then we can rewrite (\ref{equ_deltasolve}) as (\ref{equ_solve_b1rewrite}), and rewrite the left-hand side of (\ref{equ_solve_b1rewrite}) as
\been\begin{split} v(\delta):=&\frac{\rho Q\frac{\alpha(\delta+1)^2}{b}(\delta+A_0)(1-\rho+\rho Q\frac{\alpha(\delta+1)^2}{b})}{(1-\rho)(1+\rho Q\frac{\alpha(\delta+1)^2}{b})}\\
=&\frac{\rho Q\frac{\alpha(\delta+1)^2}{b}(\delta+A_0)}{1-\rho}(1-\frac{\rho}{1+\rho Q\frac{\alpha(\delta+1)^2}{b}}). \end{split}\enen

It is easy to check that $Q\frac{\alpha(\delta+1)^2}{b}$ and $1-\frac{\rho}{1+\rho Q\frac{\alpha(\delta+1)^2}{b}}$ above increase with $\delta\geq 0$. Thus, $v(\delta)$ increases with $\delta$. Note that $v(\delta=0)$ is close to $0$ for small delay $A_0$, and $v(\delta\rightarrow+\infty)=+\infty$. Since the right-hand side of (\ref{equ_solve_b1rewrite}) is a constant, the positive fixed point $\delta$ is the unique solution to (\ref{equ_solve_b1rewrite}). \qed

\subsection{$\varepsilon$-optimality for the Simplified Pricing in (\ref{equ_pt_infinity_horizon})}

We note that the approximate dynamic pricing is further simplified to (\ref{equ_pt_infinity_horizon}) by using the steady-states $Q$ in (\ref{equ_solve_Q}) and $M$ in (\ref{equ_stableM}) for infinite time horizon. It is unlike (\ref{equ_pt_finite}) which still recursively updates $Q_t$ in (\ref{equ_Qt}) and $M_t$ in (\ref{equ_Mt}) for updating $p(t)$ in finite horizon case. By using this simple dynamic price $p^{\infty}(t)$ without recursive computing over time, we wonder its performance for a finite $T$ horizon and denote the resulting expected discounted cost objective as $U^{\infty}(T)$. In the following proposition, we prove that $U^{\infty}(T)$ under simplified approximate dynamic pricing in (\ref{equ_pt_infinity_horizon}) is $\varepsilon$-optimal compared with the expected discounted cost $U(T)$ in (\ref{equ_UTobjective}) under approximate dynamic pricing $p(t)$ in (\ref{equ_pt_finite}). 

\begin{pro}\label{pro_squeeze} $\forall$ $T>0$, there always exists an $\varepsilon_T>0$ such that
\bee U(T)\leq U^{\infty}(T)\leq U(T)+\varepsilon_T, \ene
and we have $\lim_{T\rightarrow\infty}\varepsilon_T=0$ for a sufficiently large horizon $T$.
\end{pro}

\textbf{Proof:} Since $Q_t$ and $M_t$ converge to the steady-states $Q$ and $M$ respectively in the reverse time order, there exists a time threshold $t_0$ ($t_0$ time slots before the end of time horizon $T$) such that for any $t\leq T-t_0$, $Q_t=Q$ and $M_t=M$. Then, for any $t\leq T-t_0$, we have $E[A(t)]=E[A^{\infty}(t)]$ in (\ref{equ_At_forinfty}). Moreover, according to (\ref{equ_pt_finite}) and (\ref{equ_pt_infinity_horizon}), we have $p(t)=p^{\infty}(t)$ for any $t\leq T-t_0$. Therefore, the expected discounted cost in (\ref{equ_UTobjective}) under (\ref{equ_pt_finite}) can be rewritten as
\begin{align}\label{equ_proof_epsi1} U(T)=&\sum_{t=0}^{T}\rho^t(E[A(t)]^2+cp^2(t))\nonumber\\
=&\sum_{t=0}^{T-t_0}\rho^t((E[A^{\infty}(t)])^2+c(p^{\infty}(t))^2)\nonumber\\
&+\sum_{t=T-t_0+1}^{T}\rho^t(E[A(t)]^2+cp^2(t))\nonumber\\
=&\sum_{t=0}^{T-t_0}\rho^t((E[A^{\infty}(t)])^2+c(p^{\infty}(t))^2)+\varepsilon_1(T), \end{align}
where $\varepsilon_1(T)=\sum_{t=T-t_0+1}^{T}\rho^t(E[A(t)]^2+cp^2(t))$.

Note that $E[A(t)]$ in (\ref{equ_At_non_infty}) and $p(t)$ in (\ref{equ_pt_finite}) are bounded. Denote $\bar{\cU}=\max(E[A(t)]^2+cp^2(t)|t\in\{T-t_0+1,\cdots,T\})$ and $\underline{\cU}=\min(E[A(t)]^2+cp^2(t)|t\in\{T-t_0+1,\cdots,T\})$. Then, we have
\bee\label{equ_squeeze1} \varepsilon_1(T)\leq \sum_{t=T-t_0+1}^{T}\rho^t\bar{\cU}=\bar{\cU}\frac{\rho^{T-t_0+1}(1-\rho^{t_0})}{1-\rho}, \ene
and
\bee\label{equ_squeeze2} \varepsilon_1(T)\geq \sum_{t=T-t_0+1}^{T}\rho^t\underline{\cU}=\underline{\cU}\frac{\rho^{T-t_0+1}(1-\rho^{t_0})}{1-\rho}. \ene

Given the fixed threshold $t_0$ time slots before the end of the time horizon $T$, according to (\ref{equ_squeeze1}) and (\ref{equ_squeeze2}), we have $\lim_{T\rightarrow\infty}\bar{\cU}\frac{\rho^{T-t_0+1}(1-\rho^{t_0})}{1-\rho}=0$ and $\lim_{T\rightarrow\infty}\underline{\cU}\frac{\rho^{T-t_0+1}(1-\rho^{t_0})}{1-\rho}=~0$. Thus, we have
\bee \lim_{T\rightarrow\infty}\varepsilon_1(T)=0. \ene

For finite horizon with steady-states $Q, M$, by using (\ref{equ_pt_infinity_horizon}), we have \begin{align}\label{equ_proof_epsi2} U^{\infty}(T)=&\sum_{t=0}^{T-t_0}\rho^t((E[A^{\infty}(t)])^2+c(p^{\infty}(t))^2)\nonumber\\
&+\sum_{t=T-t_0+1}^{T}\rho^t((E[A^{\infty}(t)])^2+c(p^{\infty}(t))^2)\nonumber\\
=&\sum_{t=0}^{T-t_0}\rho^t((E[A^{\infty}(t)])^2+c(p^{\infty}(t))^2)+\varepsilon_2(T), \end{align}
where $\varepsilon_2(T)=\sum_{t=T-t_0+1}^{T}\rho^t((E[A^{\infty}(t)])^2+c(p^{\infty}(t))^2)$.

Similarly, we can show that $\lim_{T\rightarrow\infty}\varepsilon_2(T)=0$. Since $p(t)$ in (\ref{equ_pt_finite}) is optimal for finite horizon, we have $U(T)\leq U^{\infty}(T)$. Combine (\ref{equ_proof_epsi1}) and (\ref{equ_proof_epsi2}), we have $U^{\infty}(T)=U(T)+\varepsilon_2(T)-\varepsilon_1(T)$, and thus for $\forall$ $T>0$, there always exists a $\varepsilon_T=\varepsilon_2(T)>0$ such that $U^{\infty}(T)\leq U(T)+\varepsilon_T$ with $\lim_{T\rightarrow\infty}\varepsilon_T=0$. \qed

\begin{figure}
\centering\includegraphics[scale=0.26]{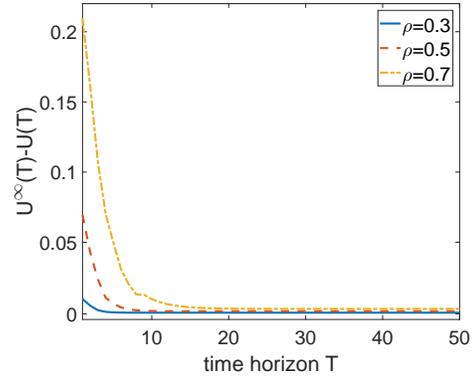}\caption{Performance gap $U^{\infty}(T)-U(T)$ versus time horizon $T$ under different discounted factor $\rho$ values when $A(0)=0, \alpha=0.9, b=2$.}\label{fig_diffU}
\end{figure}

As shown in Fig. \ref{fig_diffU}, we numerically show the difference between $U^{\infty}(T)$ and $U(T)$ reduces as $T$ increases, which approaches $0$ for sufficiently large $T$. This is consistent with this proposition above and tells that the simple pricing in (\ref{equ_pt_infinity_horizon}) performs well once $T$ is large (e.g., $T\geq 20$ in this numerical example and is not necessarily infinite). In addition, Fig.~\ref{fig_diffU} also shows that the convergence rate of (\ref{equ_pt_infinity_horizon}) to approach (\ref{equ_pt_finite}) decreases with the discounted factor $\rho$. This is because as $\rho$ increases, the impact of the ages and prices of further time slots on the total cost objective $U^{\infty}(T)$ would increase, which will result in greater cost gap.

\section{Decentralized Pricing and Mean Field Analysis for Multi-zone AoI System}\label{sec_Meanfield}





In this section, we extend the AoI control from a single zone to many self-operated zones who coordinate to provide the same type of information service (e.g., live traffic guidance) to end-customers. In reality, there are a large number $N$ of zones in a city to update the information (e.g., traffic road information), and the users arriving in each zone are invited to sense and send back real-time information to keep the overall information fresh. Due to the heterogeneity of the zones (e.g., downtown or suburban areas), we practically consider that their user arrival rates and initial AoI are different.

\subsection{Modelling of Multi-zone AoI System}

More specifically, we consider $N$ zones with heterogenous user arrival rates $\alpha_i$ and initial ages $A_i(0), i\in\{1,\cdots,N\}$, and they self-operate to decide pricing incentives locally and control their own AoI dynamics. In reality, the user arrival rates $\alpha_i$ of neighboring zones may be correlated. Denote the joint probability mass function of users' arrivals in zone $i$ and other zones $j\neq i$ as $Pr(s_i(t), s_{-i}(t))$, where $s_{-i}(t)=\{s_{1}(t),...,s_{i-1}(t),s_{i+1}(t),...,s_{N}(t)\}$, $s_i(t)=1$ indicates the user's arrival in time slot $t$ and $s_i(t)=0$ otherwise. As users' random arrivals over time in each zone are identical distributed, for each zone $i$, its user arrival rate $\alpha_i$ is given by its marginal probability $Pr(s_i(t)=1)=\sum_{j\neq i}\sum_{s_{j}(t)=0}^1Pr(s_i(t)=1, s_{-i}(t))$. For example, if there are two zones $i$ and $j$ and their user arrival rates are closely related in the joint probability of users' arrivals in both zones. Given $Pr(s_i(t)=1,s_{j}(t)=1)=0.7, Pr(s_i(t)=1,s_{j}(t)=0)=0.1, Pr(s_i(t)=0,s_{j}(t)=1)=0.1$ and $Pr(s_i(t)=0,s_{j}(t)=0)=0.1$, the user arrival rate $\alpha_i$ of zone $i$ is updated to $\alpha_i=Pr(s_i(t)=1)=Pr(s_i(t)=1,s_{j}(t)=1)+Pr(s_i(t)=1,s_{j}(t)=0)=0.8$. As end-customers are moving and may travel across various zones, each self-operated zone not only needs to care about the AoI within its own zone but also the other zones' information freshness to ensure overall service experience to end-customers \cite{li2017dynamic}. For example, Waze and Waycare share their traffic data sampled in different zones to provide the same type of live guidance service and guide the drivers to travel between different zones \cite{wazewaycare}.

In this new multi-zone self-operated system, each zone $i$ should consider a weighted sum AoI consisting of its own age and the average age of all the zones, i.e., $w_iE[A_i(t)]+(1-w_i)\frac{\sum_{j=1}^NE[A_j(t)]}{N}$ with zone $i$'s weight $w_i\in [0,1]$. A special case of this model is $w_i=1$ where each zone is not cooperative at all and only cares about its own AoI. As studied in Sections \ref{sec_approximate_dynamic} and \ref{sec_steady_infty}, then the system operation simply degenerates to independent operation by each zone, and our dynamic pricing schemes in Propositions \ref{pro_optimal_pt} and \ref{pro_steadyprice} for each zone $i$ can be applied. In the general model with $w_i<1$ in this section, however, the average information age across all the zones exhibits network externality and makes the multi-zone joint pricing design over time and zone (space) more challenging here.

Note that the weighted sum AoI of each zone $i$ is affected by all the other zones' AoI and their corresponding dynamic pricing. Different from (\ref{equ_A_dynamic})-(\ref{equ_UTobjective}), the objective of each zone $i$ is to choose its optimal dynamic pricing $p_i=\{p_i(t)|t\in\{0,\cdots,T\}\}$ as the best response to $p_{-i}=\{p_1,\dots, p_{i-1}, p_{i+1}, \dots, p_N\}$ and minimize its expected discounted cost, i.e.,
\begin{align} \bar{U}_i(p_i,p_{-i})=&\min_{p_i(t),t\in\{0,\cdots,T\}}\sum_{t=0}^{T}\rho^t\Big(\frac{\alpha_i}{b}p_i^2(t)\notag\\
&+\big(w_iE[A_i(t)]+(1-w_i)\frac{\sum_{j=1}^NE[A_j(t)]}{N}\big)^2\Big), \label{equ_Ui}\end{align}
\bee\label{equ_Ai_dynamic}\begin{split} \text{s.t.}~E[A_i(t+1)]=&E[A_i(t)]-(E[A_i(t)]-A_0)\alpha_i \frac{p_i(t)}{b}\\&+(1-\alpha_i\frac{p_i(t)}{b}), \end{split}\ene
where $\alpha_i, A_i(0)$ are different due to different user arrival rates and initial ages in different zones. Only when $w_i=1$, the multi-zone decentralized pricing problem degenerates to the single zone problem in (\ref{equ_A_dynamic})-(\ref{equ_UTobjective}).

Similar to the analysis method in Section \ref{sec_approximate_dynamic}, we first transform the nonlinear dynamics of the expected age in (\ref{equ_Ai_dynamic}) to linear dynamics, by introducing a time-average estimator $\delta_i$ for zone $i$. That is, the AoI dynamics in (\ref{equ_Ai_dynamic}) changes to
\bee\label{equ_Ai_dynamic_estimate} E[A_i(t+1)]=E[A_i(t)]-\delta_i\alpha_i \frac{p_i(t)}{b}+(1-\alpha_i\frac{p_i(t)}{b}), \ene
where the time-average term $\delta_i$ is estimated in the following way as in (\ref{equ_delta}):  \bee\label{equ_deltai} \delta_i=\frac{1-\rho}{1-\rho^T}\sum_{t=0}^{T-1}\rho^t (E[A_i(t)]-A_0). \ene
Note that $\delta_i$ is different for different zones and is affected by all the zones' user arrival rates $\alpha_j$ and ages $E[A_j(t)]$, $j\in\{1,\cdots,N\}, t\in\{0,\cdots,T-1\}$ over time due to the multi-zone coupling of weighted sum AoI in the objective (\ref{equ_Ui}).

\subsection{Analysis of Decentralized Mean Field Pricing}


In this section, we first propose mean-field term to approximate the average age dynamics of all the zones for the ease of operation at individual zones, and then analyze the decentralized mean-field pricing for each zone given any mean-field term and any zone's time-average estimator. Later in Section \ref{sec_find_fixed_mean field}, we will further analyze how to determine the mean field term and time-average estimators for finalizing pricing update.

Regarding the minimization of the discounted cost function in (\ref{equ_Ui}), the dynamic pricing design $p_i(t)$ of each zone $i$ must take into account the average age $\frac{\sum_{j=1}^NE[A_j(t)]}{N}$ of all the zones. In practice, it is not easy for each zone to keep tracking AoI dynamics of many other zones for adapting its own pricing due to large communication overhead. To make the AoI control of each self-operated zone easy to implement in practice, we desire a decentralized dynamic pricing without requiring all the zones to exchange information with each other from time to time. Therefore, we propose the following mean field term $\phi(t)$ which should satisfy
\begin{equation}\label{equ_phit}
\phi(t)=\frac{\sum_{j=1}^NE[A_j(t)]}{N}
\end{equation}
to estimate the average information age over time. Then each zone $i$ aims to design the decentralized mean field pricing $p_i(t),t\in\{0,\cdots,T\}$ according to $\phi(t)$ for minimizing its estimated total discounted cost. The problem changes from (\ref{equ_Ui})-(\ref{equ_Ai_dynamic}) to the following:
\begin{align}\label{equ_Ji} J_i(p_i,\phi)=&\min_{p_i(t),t\in\{0,\cdots,T\}}\sum_{t=0}^{T}\rho^t\Big(\frac{\alpha_i}{b}p_i^2(t)\notag\\
&+\big(w_iE[A_i(t)]+(1-w_i)\phi(t)\big)^2\Big), \end{align}
\rightline{\text{s.t.}$E[A_i(t+1)]=E[A_i(t)]-\delta_i\alpha_i \frac{p_i(t)}{b}+(1-\alpha_i\frac{p_i(t)}{b}).$ ~(\ref{equ_Ai_dynamic_estimate})}

Given any mean-field term $\phi(t)$ in (\ref{equ_phit}) and any zone $i$'s time-average estimator $\delta_i$ in (\ref{equ_deltai}), we can analyze the decentralized mean field pricing scheme for the dynamic system in (\ref{equ_Ai_dynamic_estimate}) and (\ref{equ_Ji}), by using the similar method as in Proposition \ref{pro_optimal_pt}. 



\begin{pro}\label{pro_pitmean} In the dynamic system in (\ref{equ_Ai_dynamic_estimate}) and (\ref{equ_Ji}), the optimal decentralized mean field pricing $p_i^*(t)$ of zone $i$ at time $t$ is
\begin{align}\label{equ_pt_finitei} p_i^*(t)=&\frac{\rho (\delta_i+1)M_{i,t+1}(\phi(t+1))}{2+2\rho Q_{i,t+1}\frac{\alpha_i(\delta_i+1)^2}{b}}\nonumber\\
&+\frac{2\rho(\delta_i+1)Q_{i,t+1}(E[A_i(t)]+1)}{2+2\rho Q_{i,t+1}\frac{\alpha_i(\delta_i+1)^2}{b}}, \end{align}
with $p_i^*(T)=0$, and $p_i^*(t)$ in (\ref{equ_pt_finitei}) is a function of mean field term $\phi(t+1)$, which is affected by AoI of all the other zones. The resulting expected age $E[A_i^*(t)]$ at time $t$ is
\begin{align}\label{equ_Ait_non_infty} E[A_i^*(t)]=&\prod_{j=1}^t\frac{1}{1+\rho Q_{i,j}\frac{\alpha_i(\delta_i+1)^2}{b}}A_i(0)\nonumber\\
&+\frac{2-\rho M_{i,t}(\phi(t))\frac{\alpha_i(\delta_i+1)^2}{b}}{2+2\rho Q_{i,t}\frac{\alpha_i(\delta_i+1)^2}{b}}\nonumber\\
&+\sum_{s=1}^{t-1}\bigg(\frac{2-\rho M_{i,s}(\phi(s))\frac{\alpha_i(\delta_i+1)^2}{b}}{2+2\rho Q_{i,s}\frac{\alpha_i(\delta_i+1)^2}{b}}\nonumber\\
&\times\prod_{j=s+1}^t\frac{1}{1+\rho Q_{i,j}\frac{\alpha_i(\delta_i+1)^2}{b}}\bigg), \end{align}
where $M_{i,t}(\phi(t))$ is a function of $\phi(t)$, and $Q_{i,t}, M_{i,t}(\phi(t))$ are obtained recursively:
\bee\label{equ_Qit} Q_{i,t}=w_i^2+\frac{\rho Q_{i,t+1}}{1+\rho Q_{i,t+1}\frac{\alpha_i(\delta_i+1)^2}{b}}, \ene
\begin{align}\label{equ_Mit} M_{i,t}(\phi(t))=&2w_i(1-w_i)\phi(t)\nonumber\\
&+\frac{\rho(M_{i,t+1}(\phi(t+1))+2 Q_{i,t+1})}{1+\rho Q_{i,t+1}\frac{\alpha_i(\delta_i+1)^2}{b}}, \end{align}
with $Q_{i,T}=w_i^2, M_{i,T}(\phi(T))=2w_i(1-w_i)\phi(T)$.
\end{pro}




We can see that, different from Proposition \ref{pro_optimal_pt} for a single zone, here $p_i^*(t), i\in\{1,\cdots,N\}$ in (\ref{equ_pt_finitei}) depends on the mean field term $\phi(t)$ and the weight $w_i$ of zone~$i$. Notice that the dynamic pricing $p_i^*(t)$ of zone $i$ will affect $E[A_i^*(t)]$ and thus $\phi(t)$. Thus, $p_i^*(t)$ will also affect the age $E[A_j^*(t)]$ of any other zone $j$ via $\phi(t)$.




%

\begin{algorithm}[t]
\caption{Iterative computation of fixed point estimators $\phi(t), t\in\{1,\cdots,T\}$ in (\ref{equ_phit}) and $\delta_i,i\in\{1,\dots,N\}$ in (\ref{equ_deltai})}
\begin{algorithmic}[1]


\STATE Initiate $\epsilon_1=1,\epsilon_2=1,\epsilon=0.001,j=1$, arbitrary initial $\phi^{est}_t(0)\geq 0, t\in\{1,\dots,T\}$, $\delta_i^{est}(0)\geq 0,i\in\{1,\dots,N\}$, and $\phi(t)=\phi^{est}_t(0), t\in\{1,\dots,T\}$, $\delta_i=\delta_i^{est}(0),i\in\{1,\dots,N\}$, user arrival rates $\alpha_i$ and initial ages $A_i(0), i\in\{1,\dots,N\}$

\WHILE {$\epsilon_1>\epsilon$ and $\epsilon_2>\epsilon$}

\FOR {$i=1$ to $N$}
\FOR {$t=0$ to $T-1$}
\STATE Compute $Q_{i,t}$ and $M_{i,t}(\phi(t))$ according to (\ref{equ_Qit}), (\ref{equ_Mit})
\ENDFOR
\ENDFOR

\FOR {$i=1$ to $N$}
\FOR {$t=1$ to $T$}
\STATE Compute $E[A_i^*(t)]$ according to (\ref{equ_Ait_non_infty})
\ENDFOR
\ENDFOR

\FOR {$i=1$ to $N$}
\STATE $\delta_i^{est}(j)=\frac{1-\rho}{1-\rho^T}\sum_{t=0}^{T-1}\rho^t (E[A_i^*(t)]-A_0)$, $\delta_i=\delta_i^{est}(j)$
\STATE $\epsilon_i=\delta_i^{est}(j)-\delta_i^{est}(j-1)$
\ENDFOR
\STATE $\epsilon_1=\sum_{i=1}^N\epsilon_i$
\FOR {$t=1$ to $T$}
\STATE $\phi^{est}_t(j)=\frac{\sum_{j=1}^NE[A_j^*(t)]}{N}$, $\phi(t)=\phi^{est}_t(j)$
\STATE $\epsilon'_t=\phi^{est}_t(j)-\phi^{est}_t(j-1)$
\ENDFOR
\STATE $\epsilon_2=\sum_{t=1}^T\epsilon'_t$
\STATE $j=j+1$

\ENDWHILE
\RETURN Fixed point $\phi(t),t\in\{1,\cdots,T\}$ and $\delta_i,i\in\{1,\cdots,N\}$

\end{algorithmic}
\label{alg_find_fixed_phi}
\end{algorithm}



\subsection{Update of Mean Field Term $\phi(t)$ and Time-average Estimators $\delta_i's$ for Finalizing Pricing}\label{sec_find_fixed_mean field}

Given the optimal decentralized pricing in (\ref{equ_pt_finitei}), we are ready to determine the mean field term $\phi(t)$ and time-average estimators $\delta_i's$. Note that the estimators $\delta_i's, i\in\{1,...,N\}$ to estimate the dynamic age reduction of zone $i$ and $\phi(t)$ to estimate the average age $\frac{\sum_{j=1}^NE[A_j^*(t)]}{N}$ are affected by $\{E[A_i^*(t)]|t\in\{1,...,T\},i\in\{1,...,N\}\}$ in (\ref{equ_Ait_non_infty}) of all the zones over time, which will in turn affect the age $E[A_i^*(t)]$ of each zone $i$ at time $t$. Given the user arrival rates and initial ages of all the zones, we next determine $\delta_i's, i\in\{1,...,N\}$ and $\phi(t)$ by finding the fixed points of (\ref{equ_deltai}) and (\ref{equ_phit}), respectively.


\begin{pro}\label{pro_fixedpointphi} The local estimator $\{\delta_i,i\in\{1,\cdots,N\}\}$ in (\ref{equ_deltai}) and mean field estimator $\{\phi(t), t\in\{1,\cdots,T\}\}$ in (\ref{equ_phit}) exist and are returned by Algorithm \ref{alg_find_fixed_phi}.
\end{pro}

\textbf{Proof:} For any zone $i$, substitute $\delta_i=\frac{1-\rho}{1-\rho^T}\sum_{t=0}^{T-1}\rho^t (E[A_i^*(t)]-A_0)$ and $\phi(t)=\frac{\sum_{j=1}^NE[A_j^*(t)]}{N}$ into $E[A_i^*(t)]$ in (\ref{equ_Ait_non_infty}), we can see that the age $E[A_i^*(t)]$ of zone $i$ at time $t$ is a function of the ages $\{E[A_i^*(t)]|t\in\{1,...,T\},i\in\{1,...,N\}\}$ of all the zones over time. Define the following function as a mapping from $\{E[A_i^*(t)]|t\in\{1,...,T\},i\in\{1,...,N\}\}$ to zone $i$'s age $E[A_i^*(t)]$ in (33) at time $t$:
\bee\label{equ_phit_for_diff_i}\begin{split} &\Gamma_{i,t}(\{E[A_i^*(t)]|t\in\{1,...,T\},i\in\{1,...,N\}\})=E[A_i^*(t)].  \end{split}\ene
To summarize any possible mapping $\Gamma_{i,t}$ in (\ref{equ_phit_for_diff_i}), we can define the following vector function as a mapping from $\{E[A_i^*(t)]|t\in\{1,...,T\},i\in\{1,...,N\}\}$ to the set of all the zones' ages over time:
\begin{align}\label{equ_Gamma} &\Gamma(\{E[A_i^*(t)]|t\in\{1,...,T\},i\in\{1,...,N\}\})\nonumber\\=&\Big(\Gamma_{1,1}(\{E[A_i^*(t)]|t\in\{1,...,T\},i\in\{1,...,N\}\}),\dots,\nonumber\\
&\Gamma_{1,T}(\{E[A_i^*(t)]|t\in\{1,...,T\},i\in\{1,...,N\}\}),\dots,\nonumber\\
&\Gamma_{N,1}(\{E[A_i^*(t)]|t\in\{1,...,T\},i\in\{1,...,N\}\}),\dots,\nonumber\\
&\Gamma_{N,T}(\{E[A_i^*(t)]|t\in\{1,...,T\},i\in\{1,...,N\}\})\Big). \end{align}
Thus, the fixed point to $\Gamma(\{E[A_i^*(t)]|t\in\{1,...,T\},i\in\{1,...,N\}\})=(\{E[A_i^*(t)]|t\in\{1,...,T\},i\in\{1,...,N\}\})$ in (\ref{equ_Gamma}) should be reached to let $\delta_i$ and $\phi(t)$ replicate $\frac{1-\rho}{1-\rho^T}\sum_{t=0}^{T-1}\rho^t (E[A_i^*(t)]-A_0)$ and $\frac{\sum_{j=1}^NE[A_j^*(t)]}{N}$, respectively.

We can check that $\Gamma_{i,t}\leq A_{i}(0)+t$. Define set $\Omega'=[0,A_1(0)+1]\times\cdots\times[0,A_1(0)+T]\times\cdots\times[0,A_N(0)+1]\times\cdots\times[0,A_N(0)+T]$. Since $\Gamma_{i,t}$ is continuous, $\Gamma$ is a continuous mapping from $\Omega'$ to $\Omega'$. According to the Brouwer's fixed-point theorem, $\Gamma$ has a fixed point in $\Omega'$. \qed

\begin{figure}
\centering\includegraphics[scale=0.29]{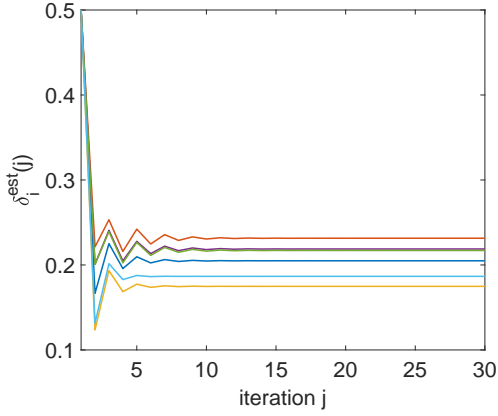}\caption{Convergence of average age estimator $\delta_i^{est}(j), i\in\{1,...,N\}$ to the fixed point $\delta_i, i\in\{1,...,N\}$ by Algorithm \ref{alg_find_fixed_phi} when $N=6$ and $T=20$.}\label{fig_aoiMFdeltaConverge}
\end{figure}

\begin{figure}
\centering\includegraphics[scale=0.29]{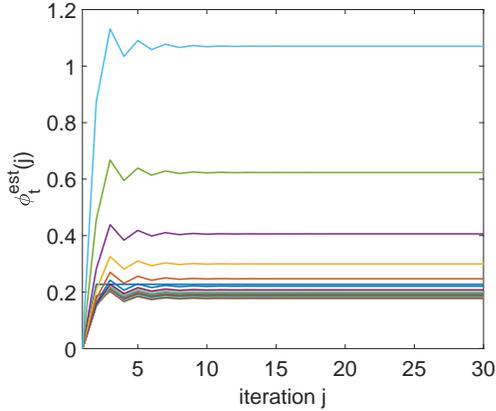}\caption{Convergence of mean field estimator $\phi_t^{est}(j), t\in\{1,...,T\}$ to the fixed point $\phi(t), t\in\{1,...,T\}$ by Algorithm \ref{alg_find_fixed_phi} when $N=6$ and $T=20$.}\label{fig_aoiMFphiConverge}
\end{figure}

Let us explain the procedure of Algorithm \ref{alg_find_fixed_phi} in the following. Given any initial estimators $\delta_i^{est}(j)$ and $\phi^{est}_t(j)$ in round $j$, we iteratively obtain the resulting expected ages $E[A_i^*(t)]$ for each zone $i$ according to (\ref{equ_Ait_non_infty}). By repeating the iteration process until $\delta_i^{est}(j+1)\rightarrow\delta_i^{est}(j)$ and $\phi^{est}_t(j+1)\rightarrow\phi^{est}_t(j)$ within arbitrarily small error $\epsilon$, the computation complexity of Algorithm \ref{alg_find_fixed_phi} is $O(\frac{TN}{\epsilon})$. Different from Algorithm \ref{alg_find_fixed_b1} for searching only in the time domain, here Algorithm \ref{alg_find_fixed_phi} jointly searches through both time and zone domains. As shown in Figs. \ref{fig_aoiMFdeltaConverge} and \ref{fig_aoiMFphiConverge}, the local estimator $\delta_i^{est}(j)$ of each zone $i\in\{1,...,N\}$ and mean field estimator $\phi_t^{est}(j)$ of each time slot $t\in\{1,...,T\}$ converge to the fixed point $\delta_i, i\in\{1,...,N\}$ and $\phi(t), t\in\{1,...,T\}$ within $20$ iterations, respectively.

\begin{figure}
\centering\includegraphics[scale=0.32]{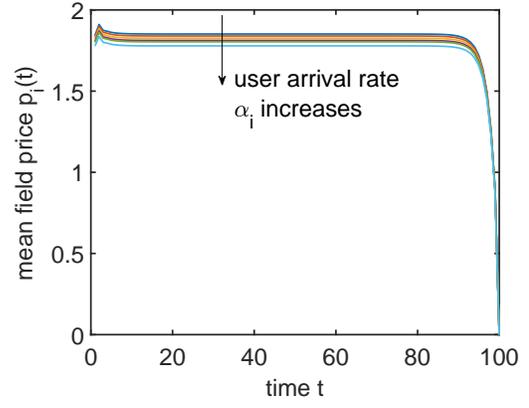}\caption{Impact of user arrival rate on the decentralized mean field pricing $p_i^*(t)$ in (\ref{equ_pt_finitei}) under original AoI dynamics (\ref{equ_Ai_dynamic}). Here we consider $N=6$ zones with randomly selected user arrival rates $\alpha_i's$.}\label{fig_meanfielduserarrivalrate}
\end{figure}

In the simulations, by applying the decentralized mean field pricing $p_i^*(t)$ in (\ref{equ_pt_finitei}) to the original nonlinear system (\ref{equ_Ai_dynamic}), we first consider $N=6$ zones with identical initial age $A_i(0)=0$ and weight $w_i=0.7$, and choose varying user arrival rate $\alpha_i$ to show the impact of user arrival rates on the mean field pricing. As shown in Fig.~\ref{fig_meanfielduserarrivalrate}, the mean field pricing $p_i^*(t)$ of each zone $i$ converges to a steady-state fast (within $10$ time slots), and a higher user arrival rate $\alpha_i$ motivates this zone to set a lower mean field pricing $p_i^*(t)$ as compared to the other zones. This is because the zone can patiently wait for a target user with low sampling cost if there are more user arrivals to sample. When close to the end of the time horizon $T=100$, the price $p_i(t)$ decreases to $0$ to save immediate sampling expense without worrying the future age.

\begin{figure}
\centering\includegraphics[scale=0.24]{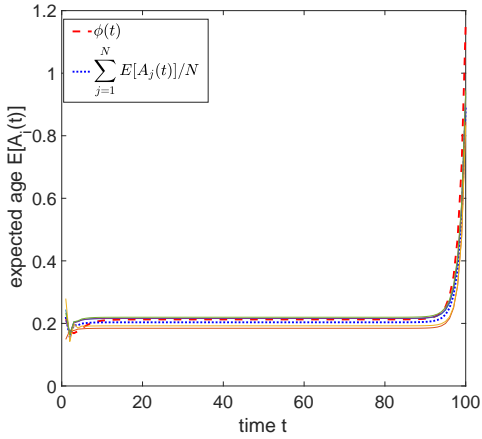}\caption{Convergence of ages $E[A_i(t)]$ for $N=20$ heterogeneous zones with different user arrival rates $\alpha_i's$, initial ages $A_i(0)'s$ and weight $w_i's$, $i\in\{1,\cdots,N\}$ under original AoI dynamics (\ref{equ_Ai_dynamic}) and decentralized mean field pricing $p_i^*(t)$ in (\ref{equ_pt_finitei}).}\label{fig_ageconsensus}
\end{figure}

In another simulation, we examine the convergence of the expected age $E[A_i(t)]$ by applying $p_i^*(t)$ in (\ref{equ_pt_finitei}) to the original nonlinear AoI dynamics (\ref{equ_Ai_dynamic}). As shown in Fig.~\ref{fig_ageconsensus}, for $N=20$ heterogeneous zones with different user arrival rates $\alpha_i's$, initial ages $A_i(0)'s$ and weight $w_i's$, $i\in\{1,\cdots,N\}$, the expected age $E[A_i(t)]$ of each zone $i$ converges to a steady-state quickly. When close to the end of the time horizon $T=100$, the expected age $E[A_i(t)]$ increases again due to decreasing $p_i(t)$ to the end. Fig.~\ref{fig_ageconsensus} also shows that our mean field term $\phi(t)$ approximates well to the average age $\frac{\sum_{j=1}^NE[A_j(t)]}{N}$ under original AoI dynamics (\ref{equ_Ai_dynamic}).


\subsection{Decentralized Pricing and Mean Field Term Estimation using Large Population Limit}\label{sec_asympMeanfield}





Our mean field term estimation in Algorithm \ref{alg_find_fixed_phi} require each zone $i$ to know any other zone $j$'s local data, i.e., initial age $A_j(0)$ and user arrival rate $\alpha_j$. Such local data are not constant all the time but change over different time periods (e.g., from off-peak to peak hours). Even each zone manages to collect such data over time, yet such local data sharing among a large number of zones (in a metropolis) introduces a large of communication overhead. In this section, we want to extend the decentralized solution developed in last section using large population limit.



We propose to employ the large population limit to determine the joint empirical distribution of the arrival rates and initial ages, and then predict the mean field term $\phi(t)$ accordingly. To be specific, when the zone population $N$ is sufficiently large, the arrival rate $\alpha_i$ and initial age $A_i(0)$ of any individual zone $i,i\in\{1,\dots,N\}$ no longer play important roles. We care about the occurrence frequency of any possible $\alpha_i$ value and any $A_i(0)$ value in the respective feasible sets $\cA$ and $\bR^+$ to determine the age evolution as well as the mean field term $\phi(t)$. Define the joint empirical CDF $F_N(\alpha,A_{ini})$ of the $N$ zones for the two dimensional sequences $\{(\alpha_i,A_i(0)), i\in\{1,\cdots,N\}\}$ with feasible sets $\alpha_i\in\cA$ and $A_i(0)\in\bR^+$ as:
\bee F_N(\alpha,A_{ini})=\frac{1}{N}\sum_{i=1}^N1_{(\alpha_i\leq \alpha, A_i(0)\leq A_{ini})}. \ene
Then, according to the joint empirical CDF $F_N(\alpha,A_{ini})$, we have
\bee\label{equ_FN} \frac{\sum_{j=1}^NE[A_j(t)]}{N}=\int_{\cA}\int_{\bR^+}A_{\alpha,A_{ini}}(t)dF_N(\alpha,A_{ini}), \ene
where $A_{\alpha,A_{ini}}(t)$ is the expected age of a typical zone at time $t$ with arrival rate $\alpha$ and initial age $A_{ini}$. 

Suppose there exists a continuous CDF $F_{\alpha,A_{ini}}(\alpha,A_{ini})$ with $\alpha\in\cA, A_{ini}\in\bR^+$ such that $\lim_{N\rightarrow\infty}F_N(\alpha,A_{ini})=F_{\alpha,A_{ini}}(\alpha,A_{ini})$ as widely adopted in large population models \cite{huang2007large}, \cite{huang2005nash}. Since we want the mean field term $\phi(t)$ to replicate $\frac{\sum_{j=1}^NE[A_j^*(t)]}{N}$ and we only know the joint cumulative distribution $F_{\alpha,A_{ini}}(\alpha,A_{ini})$ instead of realized $\alpha_j's$ and $A_j(0)'s$, we employ the large population limit to estimate $\phi(t)$, i.e.,
\bee\label{equ_phiasym} \phi(t)=\int_{\cA}\int_{\bR^+}A_{\alpha,A_{ini}}^*(t)dF_{\alpha,A_{ini}}(\alpha,A_{ini}). \ene


Then, for any given mean field estimator $\phi(t)$ in (\ref{equ_phiasym}), we can obtain the decentralized dynamic pricing $p_{i}^*(t)$ in (\ref{equ_pt_finitei}) and expected age $E[A_{i}^*(t)]$ in (\ref{equ_Ait_non_infty}) as in Proposition \ref{pro_pitmean}. Given the joint cumulative distribution $F_{\alpha,A_{ini}}(\alpha,A_{ini})$, we can also determine the fixed point mean field estimator $\phi(t), t\in\{1,\cdots,T\}$ according to Algorithm \ref{alg_find_fixed_phi}, by replacing $\phi_t^{est}(j)$ in Line $19$ of Algorithm \ref{alg_find_fixed_phi} with \bee \phi_t^{est}(j)=\int_{\cA}\int_{\bR^+}A_{\alpha,A_{ini}}^*(t)dF_{\alpha,A_{ini}}(\alpha,A_{ini}). \ene Here, we should note that in Algorithm \ref{alg_find_fixed_phi}, $Q_{\alpha,t}$ and $M_{\alpha,t}(\phi(t))$ calculated according to (\ref{equ_Qit}) and (\ref{equ_Mit}) are functions of $\alpha$, and thus the resulting age $A_{\alpha,A_{ini}}^*(t)$ is a function of $\alpha$ and $A_{ini}$.

In the following proposition, we show that the mean field term $\phi(t)$ introduced in (\ref{equ_phiasym}) is a perfect estimator of the average age $\frac{\sum_{j=1}^NE[A_j^*(t)]}{N}$, as long as the zone population $N$ is sufficiently large.

\begin{figure}
\centering\includegraphics[scale=0.24]{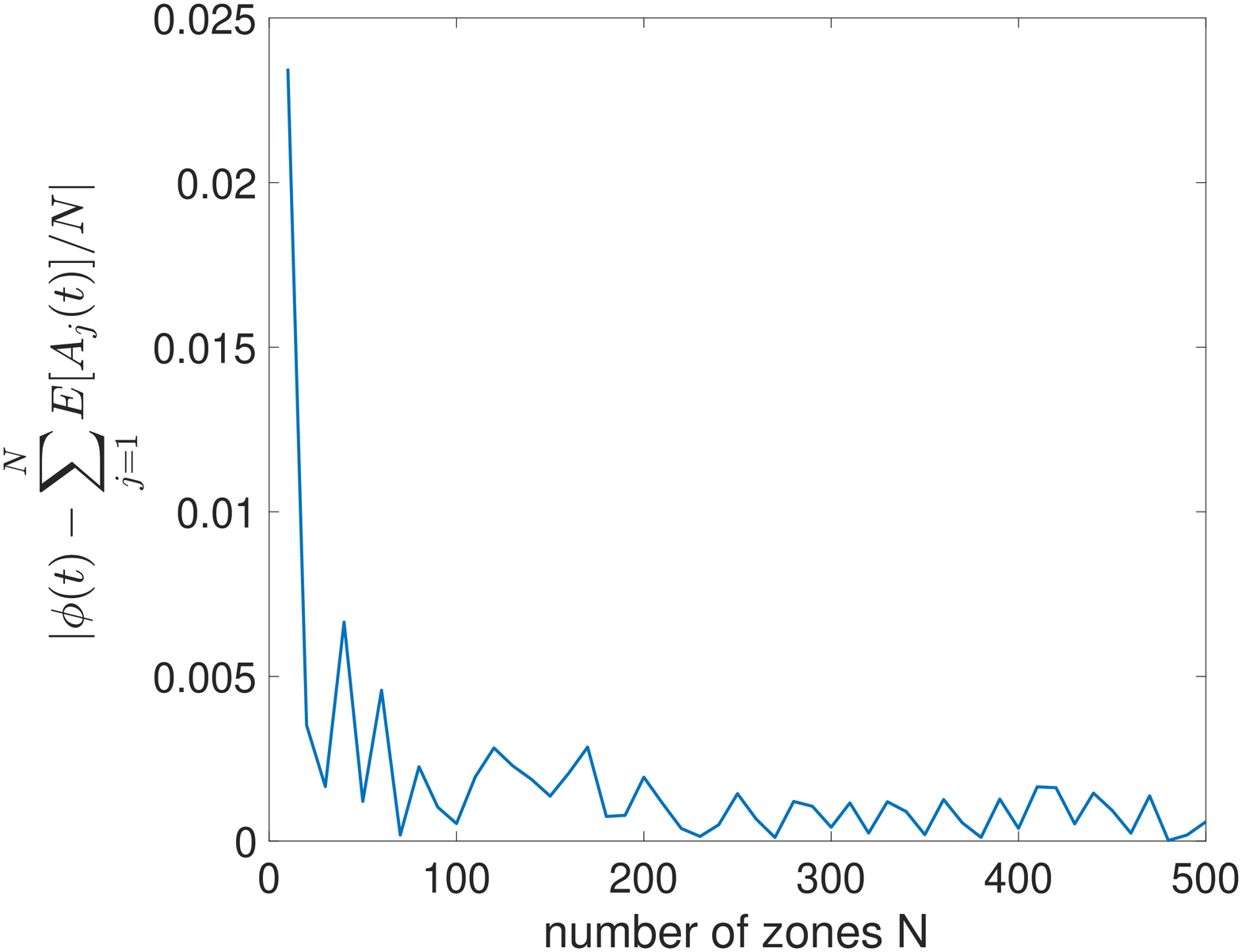}\caption{Convergence of $|\phi(t)-\frac{\sum_{j=1}^NE[A_j^*(t)]}{N}|$ with respect to the number of zones $N$ with random initial ages $A_i(0)'s$.}\label{fig_phiLargePopulation}
\end{figure}

\begin{pro}\label{pro_mu} Using the optimal price $p_i^*(t)$ in (\ref{equ_pt_finitei}) with new $\phi(t)$ in (\ref{equ_phiasym}), for any $t\geq 0$,
\bee \lim_{N\rightarrow\infty}|\phi(t)-\frac{\sum_{j=1}^NE[A_j^*(t)]}{N}|=0. \ene
\end{pro}

\textbf{Proof:} In the following proof, we first consider the one dimensional case by fixing the initial ages $A_i(0), i\in\{1,\cdots,N\}$ of all the zones to be the same. Then, the mean field term $\phi(t)$ in (\ref{equ_phiasym}) can be simplified as
\bee\label{equ_Fa} \phi(t)=\int_{\cA}A_{\alpha}^*(t)dF(\alpha),\ene
where $A_{\alpha}^*(t)$ is the expected age of the zone at time $t$ with user arrival rate $\alpha$.

According to (\ref{equ_FN}) and (\ref{equ_Fa}), we have
\bee\label{equ_phiAj}\begin{split} &|\phi(t)-\frac{\sum_{j=1}^NE[A_j^*(t)]}{N}|\\=&|\int_{\cA}A_{\alpha}^*(t)dF(\alpha)-\int_{\cA}A_{\alpha}^*(t)dF_N(\alpha)|. \end{split}\ene

Then, we extend the domain of $A_{\alpha}^*(t)$ from $\cA$ to $\bR$ in order to use Helly-Bray Theorem. Assume $\cA=[\underline{a},\overline{a}]$. Define
\begin{equation}\label{s*optimal}
A_{\alpha}'(t)=\left\{
\begin{array}{l}
A_{\alpha}^*(t), \text{for $\alpha\in\cA$;} \\
A_{\overline{a}}^*(t),  \text{for $\alpha>\overline{a}$;}\\
A_{\underline{a}}^*(t),  \text{for $\alpha<\underline{a}$.}\\
\end{array}
\right.
\end{equation}

%

Thus, we have extended function $A_{\alpha}^*(t), \alpha\in\cA$ to $A_{\alpha}'(t),\alpha\in\bR$ with $\int_{\cA}A_{\alpha}^*(t)dF(\alpha)=\int_{\bR}A_{\alpha}'(t)dF(\alpha)$ and $\int_{\cA}A_{\alpha}^*(t)dF_N(\alpha)=\int_{\bR}A_{\alpha}'(t)dF_N(\alpha)$. Since $Q_{\alpha,t}>0$ in (\ref{equ_Qit}) and $M_{\alpha,t}(\phi(t))>0$ in (\ref{equ_Mit}) are continuous in $\alpha$, and the denominators of $A_{\alpha}^*(t)$ in (\ref{equ_Ait_non_infty}) are positive, $A_{\alpha}^*(t)$ is continuous in $\alpha\in\cA$. Note that $A_{\alpha}^*(t)\leq A_{\alpha}(0)+t$ is bounded, according to Helly-Bray Theorem, if $F_N(\alpha)$ converges weakly to $F(\alpha)$,
\bee \lim_{N\rightarrow\infty}\int_{\bR}A_{\alpha}'(t)dF_N(\alpha)=\int_{\bR}A_{\alpha}'(t)dF(\alpha). \ene


Thus, 
\bee\begin{split} &\lim_{N\rightarrow\infty}|\int_{\cA}A_{\alpha}^*(t)dF(\alpha)-\int_{\cA}A_{\alpha}^*(t)dF_N(\alpha)|\\
=&\lim_{N\rightarrow\infty}|\int_{\bR}A_{\alpha}'(t)dF(\alpha)-\int_{\bR}A_{\alpha}'(t)dF_N(\alpha)|=0. \end{split}\ene
We can extend this result similarly in the two dimension case with both heterogeneous initial ages and user arrival rates. \qed

\subsubsection{Almost Sure Asymptotic Nash Equilibrium}

In the following, we will analyze the performance of the decentralized mean field pricing $p_i^*(t)$ in (\ref{equ_pt_finitei}), which is designed based on the mean field estimator $\phi(t)=\int_{\cA}\int_{\bR^+}A_{\alpha,A_{ini}}^*(t)dF_{\alpha,A}(\alpha,A_{ini})$ without requiring each zone to know the other zones' initial ages and user arrival rates. Note that such lack of information may disrupt the decentralized mean field pricing and hinder the multi-zone self-operated system to converge to a stable state. Thus, our main concern here is whether the decentralized mean field pricing scheme in (\ref{equ_pt_finitei}) (with $\phi(t)$ in (\ref{equ_phiasym}) locally estimated by each zone) ensures system stability to reach almost sure asymptotic Nash equilibrium. 

\begin{defi} A set of dynamic pricing $\{p_i^*,i\in\{1,\cdots,N\}\}$ with $p_i^*=\{p_i^*(t)|t\in\{0,\cdots,T\}\}$ is called
an almost sure asymptotic Nash equilibrium with respect to the total expected cost objective $\{\bar{U}_i,i\in\{1,\cdots,N\}\}$ in (\ref{equ_Ui}), if there exists a sequence of non-negative variables $\{\varepsilon_N, N\geq 1\}$ such that for any $i\in\{1,\cdots,N\}$,
\been \bar{U}_i(p_i^*,p_{-i}^*)\leq \inf_{p_i}\bar{U}_i(p_i,p_{-i}^*)+\varepsilon_N, \enen
where $\lim_{N\rightarrow\infty}\varepsilon_N=0$.
\end{defi}

We next prove that the system stability is ensured by our decentralized pricing $p_i^*(t)$ in (\ref{equ_pt_finitei}) and $\phi(t)$ in (\ref{equ_phiasym}).

\begin{pro}\label{pro_almostNE} The decentralized mean field pricing $\{p_i^*,i\in\{1,\dots,N\}\}$ in (\ref{equ_pt_finitei}) generated through the mean field estimator $\phi(t)$ in (\ref{equ_phiasym}) forms an almost sure asymptotic Nash equilibrium for reaching the system stability, i.e., for any $i\in\{1,\cdots,N\}$,
\bee\label{equ_thm_NE} \bar{U}_i(p_i^*,p_{-i}^*)\leq \inf_{p_i}\bar{U}_i(p_i,p_{-i}^*)+\varepsilon_N, \ene
where
\bee \varepsilon_N=\sqrt{\sum_{t=0}^{T}\rho^t(\phi(t)-\frac{\sum_{j=1}^NE[A_j^*(t)]}{N})^2}, \ene
with $\lim_{N\rightarrow\infty}\varepsilon_N=0$.
\end{pro}

\textbf{Proof:} For any $i\in\{1,\cdots,N\}$, we have
\bee\label{equ_NE_prove1}\begin{split} &\bar{U}_i(p_i^*,p_{-i}^*)\\
=&\sum_{t=0}^{T}\rho^t\Big(\big(w_iE[A_i(t)]+(1-w_i)\phi(t)\big)^2+\frac{\alpha_i}{b}(p_i^*(t))^2\Big)\\
&+\sum_{t=0}^{T}\rho^t(1-w_i)^2(\phi(t)-\frac{\sum_{j=1}^NE[A_j^*(t)]}{N})^2\\
&+2\sum_{t=0}^{T}\rho^t\Big(\big(w_i(1-w_i)E[A_i^*(t)]+(1-w_i)^2\phi(t)\big)\\
&\times(\phi(t)-\frac{\sum_{j=1}^NE[A_j^*(t)]}{N})\Big)\\
=&J_i(p_i^*,\phi)+(1-w_i)^2(\varepsilon_N)^2+\chi_N.
\end{split}\ene

By the Cauchy-Schwarz inequality, we have
\bee\begin{split} \chi_N\leq & 2\sqrt{\Big(\sum_{t=0}^{T}\rho^t\big(w_i(1-w_i)E[A_i^*(t)]+(1-w_i)^2\phi(t)\big)^2\Big)}\\
&\times\sqrt{\Big(\sum_{t=0}^{T}\rho^t(\phi(t)-\frac{\sum_{j=1}^NE[A_j^*(t)]}{N})^2\Big)}\\
= 2\varepsilon_N&\sqrt{\Big(\sum_{t=0}^{T}\rho^t\big(w_i(1-w_i)E[A_i^*(t)]+(1-w_i)^2\phi(t)\big)^2\Big)}.
\end{split}\ene
Since $\phi(t)$ is bounded due to bounded $E[A_i(t)], i\in\{1,\cdots,N\}$, $\chi_N=O(\varepsilon_N)$ and thus $(\varepsilon_N)^2+\chi_N=O(\varepsilon_N)$.

Note that $p_i^*$ is the optimal solution with respect to the estimated discounted cost function $J_i(p_i,\phi)$ in (\ref{equ_Ji}), i.e., $p_i^*=\arg\inf_{p_i}J_i(p_i,\phi)$. Then, similar to the above analysis, we have \bee\label{equ_NE_prove2} J_i(p_i^*,\phi)=\inf_{p_i}J_i(p_i,\phi)\leq  \inf_{p_i}\bar{U}_i(p_i,p_{-i}^*)+O(\varepsilon_N). \ene

Therefore, according to (\ref{equ_NE_prove1}) and (\ref{equ_NE_prove2}), we obtain (\ref{equ_thm_NE}). Further,
\bee\begin{split} \varepsilon_N=&\sqrt{\sum_{t=0}^{T}\rho^t(\phi(t)-\frac{\sum_{j=1}^NE[A_j^*(t)]}{N})^2}\\
\triangleq&\sqrt{\sum_{t=0}^{T}\rho^t(\mu_N(t))^2}<\sup_t\mu_N(t)\sqrt{\frac{1-\rho^{T+1}}{1-\rho}}. \end{split}\ene

Based on Proposition \ref{pro_mu}, for any $t>0$, $\lim_{N\rightarrow\infty}\mu_N(t)=0$. Therefore, we have $\lim_{N\rightarrow\infty}\varepsilon_N=0$. \qed

\section{Conclusion}\label{sec_conclusion}

In this paper, we have studied the AoI control by using dynamic pricing to well balance the AoI evolution and monetary payments over time. We have formulated this problem as a nonlinear constrained dynamic program under the incomplete information about users' random arrivals and their private sampling costs. To avoid the curse of dimensionality, we first propose a weighted time-average term to estimate the dynamic AoI reduction, and successfully solve the approximate dynamic pricing in closed-form. We further provide the steady-state analysis for an infinite time horizon, and show that the pricing scheme (though in closed-form) can be further simplified to an $\varepsilon$-optimal version without recursive computing over time. Finally, we extend the AoI control from a single zone to many zones with heterogeneous user arrival rates and initial ages, where each zone cares not only its own AoI dynamics but also the average AoI of all the zones to provide a holistic service. Accordingly, we design decentralized mean field pricing for each zone to self-operate by using a mean field term to estimate the average age dynamics of all the zones, which does not even require many zones to exchange their local data with each other.


\bibliographystyle{IEEEtran}
\bibliography{AoIref}

\end{document}